\documentclass[aps,prb,reprint,superscriptaddress,preprintnumbers]{revtex4-2}
\usepackage{amsfonts, amssymb, amsmath,graphicx}
\usepackage{multirow}
\usepackage{dcolumn}
\usepackage{bm}
\usepackage{color}
\usepackage[colorlinks=true,allcolors=blue]{hyperref}

\begin{document}

\preprint{RIKEN-iTHEMS-Report-23}
\title{Two-dimensional lattice with an imaginary magnetic field}

\author{Tomoki Ozawa}
\affiliation{Advanced Institute for Materials Research (WPI-AIMR), Tohoku University, Sendai 980-8577, Japan}
\author{Tomoya Hayata}
\affiliation{Departments of Physics, Keio University, 4-1-1 Hiyoshi, Kanagawa 223-8521, Japan}
\affiliation{RIKEN iTHEMS, RIKEN, Wako 351-0198, Japan}

\date{\today}

\newcommand{\tom}[1]{{\color{red} #1}}

\begin{abstract}
We introduce a two-dimensional non-Hermitian lattice model with an imaginary magnetic field and elucidate various unique features which are absent in Hermitian lattice models with real magnetic fields. To describe the imaginary magnetic field, we consider both the Landau gauge and the symmetric gauge, which are related by a generalized gauge transformation, changing not only the phase but also the amplitude of the wave function. We discuss the complex energy spectrum and the non-Hermitian Aharonov-Bohm effect as examples of properties which are due to the imaginary magnetic field independent of the generalized gauge transformation. We show that the energy spectrum does not converge as the lattice size is made larger, which comes from the intrinsic nonperiodicity of the model. However, we have found that the energy spectrum does converge if one fixes the length of one side and makes the other side longer; this asymptotic behavior can be understood in the framework of the non-Bloch band theory. We also find an analog of the Aharonov-Bohm effect; the net change of the norm of the wave function upon adiabatically forming a closed path is determined by the imaginary magnetic flux enclosed by the path, which provides an experimentally observable feature of the imaginary magnetic field.
\end{abstract}

\maketitle

\section{Introduction}
Physics of a charged particle in an external magnetic field has been of fundamental importance in condensed matter physics. In two dimensions, a charged particle in a magnetic field forms the equally spaced energy spectrum called the Landau level, which is directly responsible for phenomena such as the Landau diamagnetism, de Haas-van Alphen effect~\cite{AshcroftMermin}, and the integer and fractional quantum Hall effects~\cite{Klitzing:1980PRL,Tsui:1982PRL}. Moreover, a charged particle on a two-dimensional lattice under a magnetic field is described by the Harper-Hofstadter model~\cite{Harper:1955PPS,Hofstadter:1976PRB}, which is the paradigmatic model of the Chern insulator~\cite{Thouless:1982PRL}. 

Recently, there has been an increasing interest in non-Hermitian physics~\cite{Ashida:2020AdvPhys,Bergholtz:2021RMP}. Also in non-Hermitian quantum mechanics, the effect of vector potentials has played significant roles. For example, the seminal Hatano-Nelson model is the one-dimensional lattice model under an imaginary vector potential~\cite{Hatano:1996PRL,Hatano:1997PRB}, and has been of fundamental importance showing the non-Hermitian skin effect and nontrivial point gap topology~\cite{Gong:2018PRX,Kawabata:2019PRX,Borgnia:2020PRL,Okuma:2020PRL,Lin:2023FrontPhys}.
The non-Hermitian skin effect has been experimentally realized recently in a variety of systems including mechanical metamaterials~\cite{Brandenbourger:2019NatComm,Ghatak:2020PNAS}, electrical circuits~\cite{Helbig:2020NatPhys,Hofmann:2020PRR}, photonics~\cite{Weidemann:2020Science,Xiao:2020NatPhys,Xiao:2021PRL}, and ultracold atomic gases~\cite{Liang:2022PRL}.
The imaginary vector potential has also been crucial in understanding the Landau-Zener transition~\cite{PhysRevB.58.16051,PhysRevB.81.033103}.

With the recent experimental development of non-Hermitian quantum mechanics, one can now realize a variety of non-Hermitian models under control, and there is an increasing interest in experimentally realizing two or higher dimensional non-Hermitian models~\cite{Yamauchi:2020arXiv,Zou:2021NatComm,Palacios:2021NatComm,Shang:2022AdvSci}.
Despite this rapid progress in non-Hermitian physics and the important role of magnetic fields played in condensed matter physics, there has been little study on properties of the imaginary magnetic fields in two dimensions, namely, properties of non-Hermitian lattice systems where magnetic fields are imaginary, analogous to the imaginary vector potential in the Hatano-Nelson model.

In this paper, we explore basic properties of two-dimensional lattices with an imaginary magnetic field. We first elucidate the meaning of gauge invariance in non-Hermitian settings, to distinguish between properties which are gauge independent or gauge invariant, i.e., intrinsically due to the imaginary magnetic field, and which are gauge dependent, i.e., dependent on specific realizations and setups. We find that certain spectral properties are gauge invariant, and discuss that the asymptotic energy spectrum as one makes the length of the system larger, fixing the other length, can be nicely understood within the framework of the non-Bloch band theory.
Especially, even though the system is non-Hermitian, the asymptotic spectrum under open boundary conditions can be related to the spectrum under the periodic boundary condition satisfying certain conditions.
We also find an analog of the Aharonov-Bohm effect for imaginary magnetic fields. Upon making a wave packet move to form a closed trajectory in real space, the overall change of the norm of the wave function is related to the imaginary magnetic flux enclosed by the trajectory. Our paper lays a foundation to understand gauge invariant properties in the setup of imaginary magnetic fields, generalizing the concept of magnetic fields to two-dimensional non-Hermitian settings.

The structure of the paper is the following. In Sec.~\ref{sec:model}, we introduce the model we study, which is the two-dimensional non-Hermitian lattice model with an imaginary magnetic field. In Sec.~\ref{sec:gaugetransformation}, we discuss generalized gauge transformations which are relevant to discussing properties intrisically due to the imaginary magnetic field. In Sec.~\ref{sec:spectrum}, we discuss various properties of the complex energy spectrum of the model. We first discuss open boundary conditions in both directions and then the effect of taking one direction to be periodic, which we call the cylindrical configuration. In Sec.~\ref{sec:asym}, we discuss the asymptotic limit of the energy spectrum when we fix the length of one side and make the length of the other side infinitely long. We analyze the asymptotic spectrum using the non-Bloch band theory. In Sec.~\ref{sec:AB}, we discuss the analog of the Aharonov-Bohm effect under the imaginary magnetic field. We finally give the conclusion and future prospects in Sec.~\ref{sec:concl}.

\section{Model}
\label{sec:model}
We consider a two-dimensional square lattice with an imaginary magnetic field. We label the lattice sites by coordinates $(x,y)$, where $x$ and $y$ are both integers. We let $\psi_{x,y}$ denote the amplitude of the wave function at site $(x,y)$. The Schr\"odinger equation governing the dynamics of the system is
\begin{align}
    i\frac{d \psi_{x,y}}{dt}
    =&
    J\left( e^{i\theta_X (x-1,y)}\psi_{x-1,y} + e^{-i\theta_X (x,y)}\psi_{x+1,y} \right.
    \notag \\
    &\left. + e^{i\theta_Y (x,y-1)}\psi_{x,y-1} + e^{-i\theta_Y (x,y)}\psi_{x,y+1}\right),
\end{align}
where $t$, and $J$ are the time, and hopping parameter, respectively.
In this paper, we consider two gauge choices: the Landau gauge and the symmetric gauge. 
The Landau gauge is defined by $(\theta_X, \theta_Y) = (0, Bx)$, and the symmetric gauge is defined by $(\theta_X, \theta_Y) = (-By/2, Bx/2)$. When $B$ is real, these gauges correspond to the ordinary Landau and symmetric gauges with a real magnetic field. In this paper, however, we take $B$ to represent a purely imaginary magnetic field, $B = i\mathcal{B}$, with $\mathcal{B}$ being a real number. We note that, when $B$ is imaginary, the factors $e^{i\theta_X (x,y)}$ and $e^{i\theta_Y (x,y)}$ can have a modulus different from one, implying non-Hermiticity. 

\section{Gauge transformation}
\label{sec:gaugetransformation}
Landau and symmetric gauges are equivalent for a real magnetic field because they are related by a gauge transformation. We first review the gauge transformation in Hermitian setups, and then extend the concept to non-Hermitian settings. The gauge transformation is to consider a state which is related to the original state by a position-dependent phase factor $\psi^\prime_{x,y} = e^{i\chi (x,y)}\psi_{x,y}$, where $\chi (x,y)$ is a real function. This transformation amounts to applying a local unitary transformation to wave function. If we take the Hamiltonian $H$ in the Hermitian Landau gauge, by choosing the gauge transformation $e^{i \chi (x,y)} = e^{iBxy/2}$, the Hamiltonian $H^\prime$ transformed under the gauge transformation is in the Hermitian symmetric gauge. 
Since the gauge transformation is a unitary transformation, the energy spectrum is invariant under the gauge transformation.

Now we extend the concept of gauge transformations to non-Hermitian setups of imaginary magnetic fields. An important feature of imaginary magnetic fields is that the Landau gauge and the symmetric gauge cannot be connected via an ordinary gauge transformation $e^{i\chi (x,y)}$ determined by a real function $\chi (x,y)$. Instead, it is more appropriate to consider a generalized gauge transformation, $\psi^\prime_{x,y} = f(x,y)\psi_{x,y}$ with $f(x,y)$ being a nonzero complex function, which does not just multiply a phase factor but also allows scale change for the wavefunction. The Hamiltonian changes under this generalized gauge transformation, not by a unitary transformation but by a local (diagonal) similarity transformation. The Landau and symmetric gauges are related via the generalized gauge transformation $f(x,y) = e^{iBxy/2}= e^{-\mathcal{B}xy/2}$.
Since this generalized gauge transformation is a similarity transformation, the energy spectrum is invariant. Furthermore, upon the generalized gauge transformation, the product of hopping amplitudes as one goes around a plaquette of the square lattice does not change, implying that the imaginary magnetic field is also invariant. 

There are various realizations of non-Hermitian Hamiltonians, and what is observable depends on the individual system that one works on.
Upon studying properties of imaginary magnetic fields, one should thus make a clear distinction between what are universal properties of imaginary magnetic fields and what are gauge- and system-specific features which depend on particular realizations. We consider properties intrinsic to imaginary magnetic fields to be those invariant under the generalized gauge transformation.

\section{Energy spectrum}
\label{sec:spectrum}
As is well-known in the study of non-Hermitian Hamiltonians, an energy spectrum under periodic and open boundary conditions can take drastically different values~\cite{Hatano:1996PRL,Hatano:1997PRB,Yao:2018PRL,Okuma:2020PRL}. We should therefore analyze the energy spectrum together with the boundary conditions. We first note that, unlike the case of real magnetic fields, lattice models with imaginary magnetic fields cannot be made periodic in both $x$ and $y$ directions. For the Landau gauge, we can make the lattice periodic in the $y$ direction, but not in the $x$ direction, and for the symmetric gauge we cannot make the Hamiltonian periodic in either direction. In this paper, we call the Landau gauge with the periodic boundary condition in the $y$ direction a \textit{cylindrical configuration}. As we see, the energy spectrum under open boundary conditions in both directions and that under the cylindrical configuration are qualitatively different.

\subsection{Open boundary conditions}

We first consider the open boundary conditions.
As we have seen, the energy spectrum under the Landau and symmetric gauges are the same because they are related by the generalized gauge transformation. We also note that the energy spectrum is invariant upon the change of the origin of the coordinate: the spectrum is invariant under changing $x$ to $x + x_0$ and $y$ to $y+y_0$ in the hopping factors $\theta_X$ and $\theta_Y$. This invariance can be shown, for example, for the symmetric gauge, by noting that the shift $x \to x+x_0$ can be realized by $f(x,y) = e^{-\mathcal{B} x_0 y/2}$ and the shift $y \to y+y_0$ by $f(x,y) = e^{-\mathcal{B} x y_0/2}$. Since we do not want the imaginary magnetic fields and their properties to depend on the origin of the coordinates, these transformation properties are desirable. 

\begin{figure*}[htbp]
    \centering
    \includegraphics[width=0.95\textwidth]{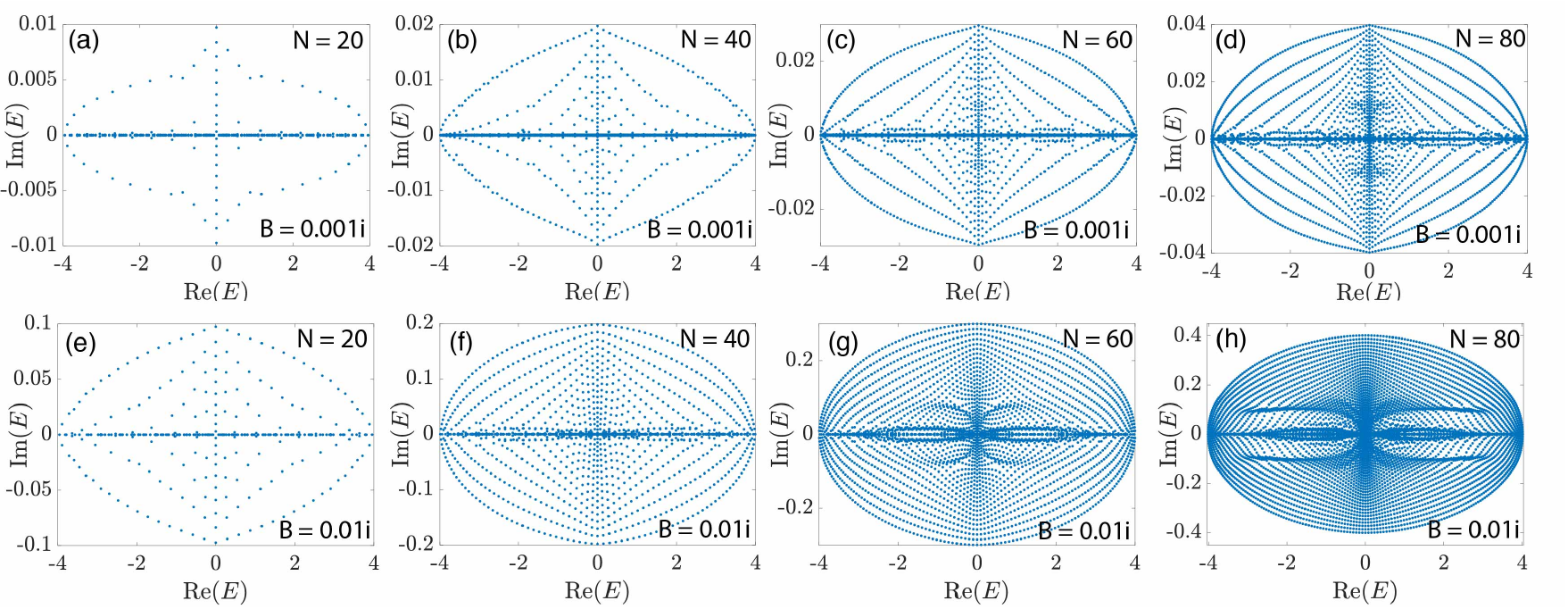}
    \caption{Energy spectrum for different system sizes and magnetic fields under open boundary conditions for (a)-(d) $B = 0.001i$ and (e)-(h) $B = 0.01i$. The system sizes are (a), (e) $N_x = N_y = 20$; (b), (f) $N_x = N_y = 40$; (c), (g) $N_x = N_y = 60$; and (d), (h) $N_x = N_y = 80$, respectively. All axes are in units of $J$.}
    \label{fig:app1}
\end{figure*}

In Fig.~\ref{fig:app1}, we plot the energy spectrum in the complex plane for a lattice of size $N_x \times N_y$ with $N_x = N_y = 20, 40, 60$, and 80 for the imaginary magnetic fields of $B = 0.001i$, and $0.01i$.
We observe that the the spread of the energy spectrum along the real axis is from $-4J$ to $4J$, whereas the spread of the energy along imaginary axis varies with the strength of the imaginary magnetic fields as well as the size of the system. This spread of the energy spectrum along real and imaginary axes can be explained assuming $B = \mathcal{B}i$ is small, as we shall show.

To understand the spectral range of $H(B)$, we expand the Hamiltonian $H(B)$ in the Taylor series around $B = 0$ and keep up to the first order in $B$ assuming $B$ is small. We write this expansion as $H(B) = H(0) + B \delta H$.
When $B$ is purely imaginary, the first-order term $B \delta H = i\mathcal{B}\delta H$ is an antisymmetric matrix. Since the zeroth order term  $H(0)$ is a symmetric matrix, $H(B) = H(0) + i\mathcal{B} \delta H$ is a decomposition of $H(B)$ into its symmetric and antisymmetric components. Since $H(0)$ is a symmetric matrix, its eigenvalues are all purely real, whereas the eigenvalues of $i\mathcal{B} \delta H$, which is an anti-symmetric matrix, are all purely imaginary. In the Appendix, we show that, for any diagonalizable real matrix $M$, by writing $M = S + A$, where $S$ and $A$ are symmetric and antisymmetric real matrices, the real part of the eigenvalues of $M$ are contained within the minimum and maximum values of the eigenvalues of $S$, and the imaginary part of the eigenvalues of $M$ are contained within the minimum and maximum imaginary eigenvalues of $A$.
The spectral range of $H(B) = H(0) + i\mathcal{B} \delta H$ is thus contained, in the real-axis direction, by the minimum and maximum eigenvalues of $H(0)$ and is contained in the imaginary-axis direction by the minimum and maximum imaginary eigenvalues of $i\mathcal{B} \delta H$.
Now, $H(0)$ is the Hamiltonian of a two-dimensional square lattice without any magnetic field. The eigenvalues in the thermodynamic limit are given by the simple sinusoidal band structure $2J \left( \cos (k_x) + \cos (k_y) \right)$, where $(k_x, k_y)$ is the two-dimensional quasimomentum. Therefore, their maximum and minimum are given by $\pm 4J$. The antisymmetric part, $i\mathcal{B} \delta H$, in the Landau gauge, is the Hamiltonian of decoupled $N_x$ one-dimensional chains with nonreciprocal hoppings; for a chain at position $x$, the hopping along the chain is $\pm \mathcal{B}x J$, where $x$ runs from some initial value $x_0$ to $x_0 + N_x-1$. (Note that, in $i\mathcal{B} \delta H$, there is no hopping between chains.)
For a chain at position $x$, its eigenvalues, as a function of the quasimomentum $k_y$ along $y$ direction, are $2iJ\mathcal{B}x \cos (k_y)$.
The range of eigenvalues of $i\mathcal{B} \delta H$ thus depends on how $x_0$ is chosen. 
To obtain the tightest bound, we can take $x_0$ to be $-(N_x-1)/2$, so $x$ takes values from $-(N_x-1)/2$ to $(N_x-1)/2$. The minimum and maximum imaginary eigenvalues of $i\mathcal{B} \delta H$ are then $\pm J(N_x - 1)\mathcal{B}i$. The spread of the energy spectrum of $H(B)$ obtained numerically in Fig.~\ref{fig:app1} along the imaginary direction agrees with the bound $\pm J(N_x - 1)\mathcal{B}i$; we note that the actual spectral spread along the imaginary direction is roughly half compared to the bound $\pm J(N_x - 1)\mathcal{B}i$. The spectrum of $i\mathcal{B} \delta H$ provides only the upper and lower bounds, and does not necessarily give limits that can be saturated.

We note that, while the spread along the real axis obtained from the eigenvalues of $H(0)$ does not depend on the system size, the spread along the imaginary axis obtained from the eigenvalues of $i\mathcal{B} \delta H$ depends linearly on $N_x = N_y = N$. Numerically calculated eigenvalues in Fig.~\ref{fig:app1} also shows that, by fixing the magnetic field and increasing the system size, the energy spectrum increases its spreading along the imaginary direction and does not converge. This nonconvergence of the energy spectrum as $N \to \infty$ is in stark contrast to ordinary two-dimensional Hermitian lattices in which increasing the system size makes the energy spectrum converge to a continuous band structure, except for possible edge-localized states. The origin of the non-convergence of our energy spectrum is because, even though the imaginary magnetic field is fixed and constant over the entire lattice, the hopping strength such as $e^{-\mathcal{B}x}$ keeps increasing in the $x$ direction.

The spectral bounds we have obtained assume that the imaginary magnetic field $B = i\mathcal{B}$ is small. As the magnetic field becomes larger, the spread of the energy spectrum of $H(B)$ can go beyond the bounds obtained by assuming small $B$. 
Even in such a case, the spectral bounds can be obtained by decomposing $H(B)$ into the symmetric and asymmetric parts.
This predicts that the bounds on both the real and imaginary parts grow exponentially as the imaginary magnetic field or the system size increases.
However, the numerical diagonalization of $H(B)$ becomes more and more difficult as $B$ becomes larger because of the existence of matrix elements which significantly differ in their magnitudes. In this paper, we focus our analysis on the case of small $B$, where we can make definite statements, both numerically and analytically, about the spectral structure and bounds.

Although the energy spectrum does not converge when both sides of the lattice are taken to increase, we will see later that the energy spectrum does converge when we fix the length of one side and take the other direction to increase. We discuss the asymptotic spectrum when one side is made longer in Sec.~\ref{sec:asym}.

\subsection{Cylindrical configuration}

\begin{figure}[htbp]
    \centering
    \includegraphics[width=0.5\textwidth]{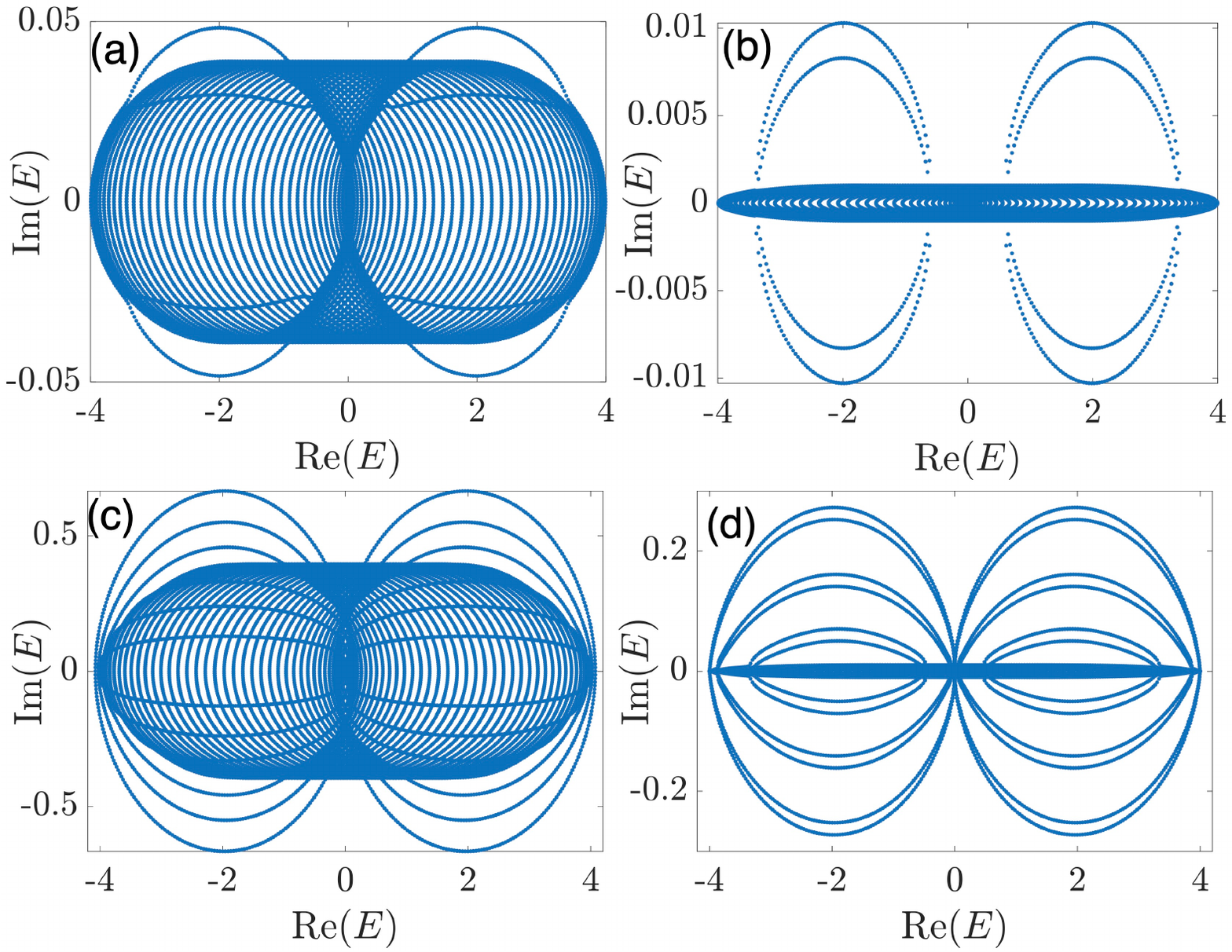}
    \caption{Energy spectrum under the periodic boundary condition along the $y$ direction (cylindrical configuration) with $N_x = 40$ for (a), (b) $B = 0.001i$ and (c), (d) $B = 0.01i$. The $x$ coordinates are labeled as $x = 0, 1, 2, \cdots, 40$ for (a), (c) and $x = -20, -19, \cdots, 19$ for (b), (d). All the axes are in units of $J$.}
    \label{fig:first}
\end{figure}

We next consider the case of the periodic boundary condition along the $y$ direction under the Landau gauge, namely, the cylindrical configuration.
The energy spectrum can be obtained by performing the Fourier transformation in the $y$ direction and diagonalizing the Hamiltonian for each momentum separately. Writing $\psi_{x,y} = \psi_x e^{ik}$, the equation to solve is
\begin{align}
    E \psi_x = J \left\{ \psi_{x-1} + \psi_{x+1} + \left( e^{-x\mathcal{B}-ik} + e^{x\mathcal{B}+ik} \right) \psi_x \right\}. \label{eq:nonblocheig}
\end{align}
We note that this is an analog of the Harper equation for the imaginary magnetic field~\cite{Harper:1955PPS}.
In Fig.~\ref{fig:first}, we plot the energy spectrum in the cylindrical configuration with two different values of the imaginary magnetic field and two different ways to choose the origin of $x = 0$. 
We find an unexpected feature that the energy spectrum depends on the origin of the coordinates. Under open boundary conditions, we saw that shifting of $x \to x + x_0$ is achieved by the generalized gauge transformation of $f(x,y) = e^{-\mathcal{B}x_0 y}$. However, this gauge transformation is not periodic in the $y$ direction and thus is not compatible with the cylindrical configuration. The energy spectrum therefore depends on how the origin of $x$ is taken. This is a unique feature of the imaginary magnetic field; the energy spectrum of the real magnetic field is independent of how the origin of $x$ is taken, even when the cylindrical configuration is taken along the $y$ direction under the Landau gauge.

\section{Asymptotic spectrum}
\label{sec:asym}
\begin{figure}
    \centering
    \includegraphics[width=0.5\textwidth]{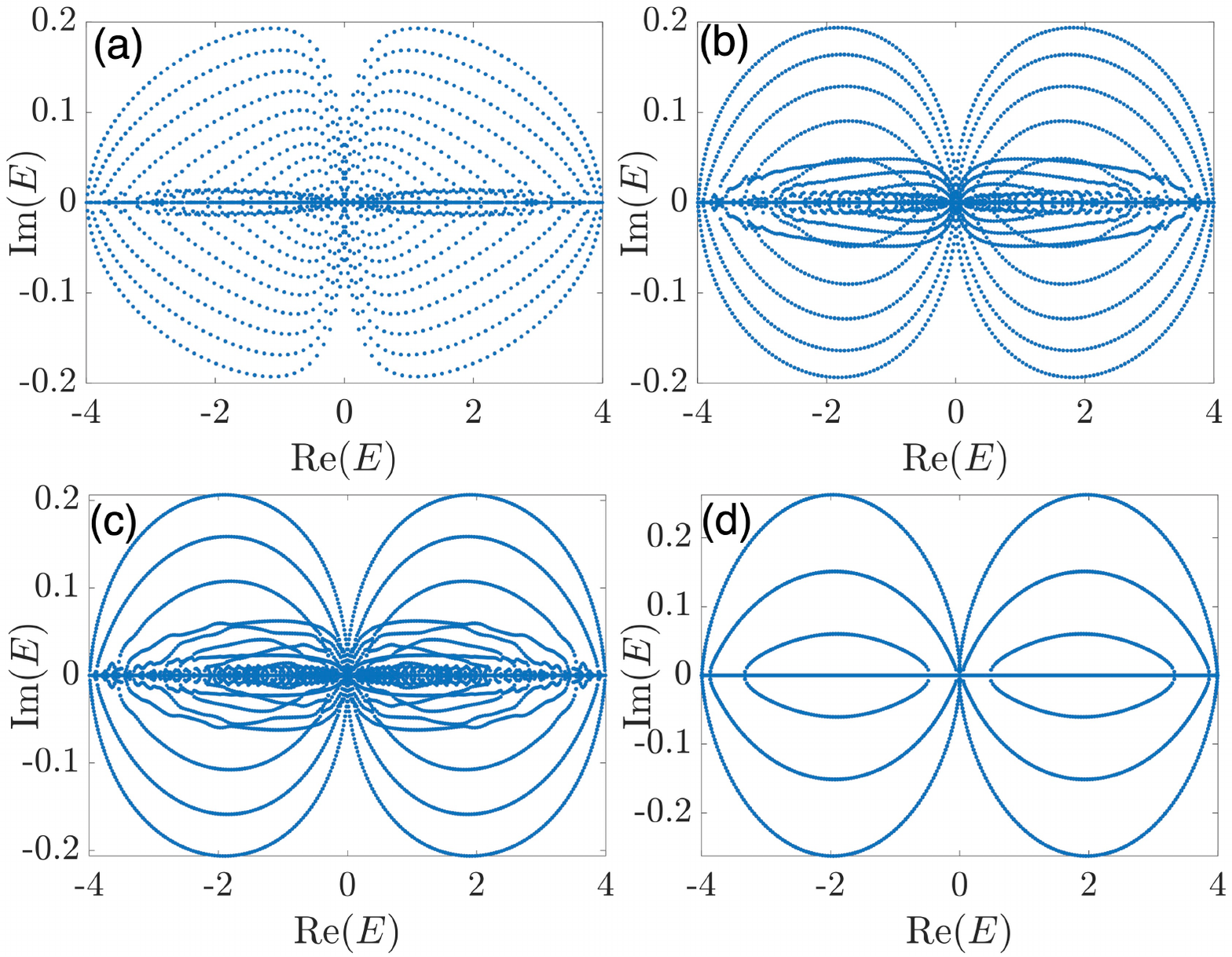}
    \caption{(a)-(c) Asymptotic energy spectrum fixing $N_x = 40$ under open boundary conditions for $B = 0.01i$. (a) $N_y = 50$. (b) $N_y = 100$. (c) $N_y = 150$. (d) Asymptotic energy spectrum predicted from the non-Bloch band theory. All axes are in units of $J$.}
    \label{fig:third}
\end{figure}

Even though the spectrum does not converge keeping $N_x = N_y$, we find that the spectrum does converge as one fixes the size of one side and makes the other side become longer.
In this section, we discuss the asymptotic behavior of the energy spectrum as one side is made longer.
In Fig.~\ref{fig:third}, we plot the energy spectrum under open boundary conditions when $B = 0.01i$, fixing $N_x = 40$ and choosing $N_y = 50$, 100, 150. One sees that the overall shape tends to stabilize as $N_y$ becomes large. We can understand this asymptotic energy spectrum in the limit of large $N_y$ by means of the non-Bloch band theory~\cite{Yokomizo:2019PRL,Yokomizo:2023PRB}. The non-Bloch band theory is a formalism to obtain the continuous energy spectrum of non-Hermitian systems under open boundary conditions. To understand the asymptotic behavior of fixing $N_x$ and making $N_y \to \infty$, we now regard the index $x$ to be an internal index of a one-dimensional system elongated along the $y$ direction. 

\subsection{Non-Bloch band theory}

To apply the non-Bloch band theory, we perform the Fourier transformation along the $y$ direction, as done in Eq.~(\ref{eq:nonblocheig}) above. In the non-Bloch band theory, we replace $e^{ik}$ by a general complex number $\beta$, and solve the above eigenvalue equation for a given value of $E$. Writing the above eigenvalue equation as $E\vec{\psi}_X = H_X(\beta) \vec{\psi}_X$, where $\vec{\psi}_X$ is a vector whose element is $\psi_x$, solutions to the eigenvalue equation for a given value of $E$ are given by the solutions of $\det [H_X(\beta) - E] = 0$.
This equation is an algebraic equation for $\beta$ with degree $2N_x$, and thus we generally have $2N_x$ solutions of $\beta$. Writing $2N_x$ solutions of $\beta$ in the ascending order of their magnitudes and labeling them as $\beta_1$, $\beta_2$, $\cdots$, the eigenvalue $E$ belongs to a continuum of energy band if and only if $|\beta_{N_x}| = |\beta_{N_x+1}|$~\cite{Yokomizo:2019PRL}. The corresponding values of $\beta_{N_x}$ and $\beta_{N_x+1}$ form the generalized Brillouin zone in the complex plane. 

As we show below, the generalized Brillouin zone coincides with the ordinary Brillouin, namely, $\beta = e^{ik}$ for real $k$, when the $x$ coordinate is labeled so $x = 0$ is in the center of the system.
We want to look for the generalized Brillouin zone and the continuum of energy bands by solving $\det [H_X(\beta) - E] = 0$.
Remembering that the spectrum under open boundary conditions is independent of the choice of the origin of $x = 0$, we can choose the coordinates most convenient for our purpose. It turns out to be particularly useful to take a coordinate system where $x = 0$ is in the center, namely, if $N_x$ is an odd number, we write $N_x = 2p + 1$ with a positive integer $p$ and take the $x$ coordinate to be $x = -p, -p + 1, \cdots, -1, 0, 1, \cdots, p-1, p$. If $N_x$ is an even number, we write $N_x = 2p$ with a positive integer $p$ and take $x = -p + \frac{1}{2}, \cdots, -\frac{3}{2}, -\frac{1}{2}, \frac{1}{2}, \frac{3}{2}, \cdots, p - \frac{1}{2}$.

Let us first consider the case where $N_x$ is an odd number: $N_x = 2p + 1$. We take the $x$ coordinate label as above so $x = 0$ is in the center. Then,
\begin{widetext}
\begin{align}
    H_X(\beta)
    =
    \begin{pmatrix}
        e^{-p\mathcal{B}}\beta + e^{p\mathcal{B}}\frac{1}{\beta} & 1 \\
        1 & e^{(-p+1)\mathcal{B}}\beta + e^{(p-1)\mathcal{B}}\frac{1}{\beta} & \ddots \\
        & \ddots & \ddots \\
        & & & e^{-\mathcal{B}}\beta + e^{\mathcal{B}\frac{1}{\beta}} & 1 \\
        & & & 1 & \beta + \frac{1}{\beta} & 1\\
        & & & & 1 & e^{\mathcal{B}}\beta + e^{-\mathcal{B}}\frac{1}{\beta} & \ddots \\
        & & & & & \ddots & \ddots & 1\\
        & & & & & & 1 & e^{p\mathcal{B}}\beta + e^{-p\mathcal{B}}\frac{1}{\beta}
    \end{pmatrix} ,
\end{align}
\end{widetext}
in units of $J$.
One thus sees that if $\beta$ is a solution of $\det [H_X(\beta) - E] = 0$, so is $1/\beta$, namely, $\det [H_X(1/\beta) - E] = 0$.
The same holds for even $N_x$.

This implies that, upon writing $2N_x$ solutions of $\det [H_X(\beta) - E] = 0$ in ascending order, we should have $\beta_{N_x + 1} = 1/\beta_{N_x}$. From this condition, together with the condition that $|\beta_{N_x}| = |\beta_{N_x + 1}|$ for $E$ to be in the continuum of energy bands~\cite{Yokomizo:2019PRL}, we arrive at the condition $|\beta_{N_x}| = |\beta_{N_x + 1}| = 1$. This implies that $\beta$ belonging to the generalized Brillouin zone can be written as $\beta = e^{ik}$ with a real number $k$, namely, the generalized Brillouin zone coincides with the ordinary Brillouin zone with quasimomentum $0 \le k < 2\pi$. 

Conversely, for a given value of $\beta = e^{ik}$ with a real $k$, all $N_x$ eigenvalues of $H_X(e^{ik})$ belong to the continuum of energy bands. This is because any eigenvalue $E$ of $H_X(e^{ik})$ satisfies $\det [H_X(e^{ik}) - E] = 0$, and when writing $2N_x$ solutions of $\beta$ in an ascending order, $\beta = e^{\pm ik}$ must appear at $N_x$th and $N_x+1$th positions. We have thus shown that the continuum of energy bands obtained upon fixing $N_x$ and making $N_y$ large is nothing but the energy spectrum in the cylindrical configuration where coordinates in the $x$ direction are taken so $x = 0$ is placed at the center.

This implies that the solutions of Eq.~(\ref{eq:nonblocheig}) for real $k$, which are nothing but the energy spectrum of the cylindrical configuration, are the asymptotic spectrum when fixing $N_x$ and making $N_y$ large under open boundary conditions. 
The fact that the generalized Brillouin zone coincides with the ordinary Brillouin zone implies that there is no non-Hermitian skin effect. This absence of the non-Hermitian skin effect is related to the $\mathcal{PT}$-symmetry present in the system~\cite{Yi:2020PRL}.

In Fig.~\ref{fig:third}(d), we show the continuum bands obtained from the energy spectrum of a cylindrical configuration, taking $x = 0$ to be at the center. We see that the spectra in Figs.~\ref{fig:third}(a)-\ref{fig:third}(c) indeed approaches that of Fig.~\ref{fig:third}(d).
With different values of $\mathcal{B}$ and $N_x$, we find that there is a general structure of a continuous spectrum along the real axis and several oval structures spread along the imaginary direction, but the exact number of ovals and the spread along the imaginary direction depend on specific values of the parameters. 
We stress that the energy spectrum under open boundary conditions does not depend on how the coordinates are chosen. Nevertheless, the asymptotic spectrum coincides with the energy spectrum in the cylindrical configuration where the coordinates are chosen in a symmetric manner.

\section{Aharonov-Bohm effect for imaginary magnetic fields}
\label{sec:AB}
We now discuss an effect analogous to the Aharonov-Bohm effect~\cite{Aharonov:1959PR} for imaginary magnetic fields.
The non-Hermitian Aharonov-Bohm effect in a parameter space due to the complex Berry phase has been experimentally observed for synthetic mechanical metamaterials~\cite{Anandwade:2023PRA,Singhal:2023PRR}, but, to our knowledge, it has never been observed in real space.
We first formulate the non-Hermitian Aharonov-Bohm effect, which is the amplification and/or decay of the magnitude of the wave function due to the imaginary magnetic flux enclosed by a closed trajectory.
Following the standard argument of the Aharonov-Bohm effect in Hermitian systems~\cite{GriffitshBook}, the wave function $\vec{\psi}$ of a wave packet, which we assume to be well localized in real space, is transported over a path $\mathcal{C}$ in the presence of a vector potential.
Then, the wave function $\vec{\psi}$ is related to the wave function $\vec{\psi_0}$ transported over the same path $\mathcal{C}$ but in the absence of a vector potential by
\begin{align}
    \vec{\psi} = \exp \left( i \int_{\mathbf{r}_i, \mathcal{C}}^{\mathbf{r}_f} \mathbf{A}(\mathbf{r}^\prime)\cdot d\mathbf{r}^\prime\right) \vec{\psi_0},
\end{align}
where the line integral is taken along the particular path $\mathcal{C}$, connecting its initial position $\mathbf{r}_i$ and the final position $\mathbf{r}_f$ of the wave packet, has been transported. 
This formula is also valid in non-Hermitian systems, where $\mathbf{A}(\mathbf{r})$ becomes a complex number.
Taking a closed path $\mathcal{C}$, the wave function after the transportation is related to the wavefunction in the absence of a magnetic field by
\begin{align}
    \vec{\psi} = \exp \left( i \oint_{\mathcal{C}} \mathbf{A}(\mathbf{r}^\prime)\cdot d\mathbf{r}^\prime\right) \vec{\psi_0}
    =
    e^{i \Phi} \vec{\psi_0},
\end{align}
where $\Phi$ is the total flux through a surface that the path $\mathcal{C}$ encloses. In our case, since the flux is pure-imaginary, the prefactor $e^{i \Phi}$ is a real number, giving rise to the amplifiaction and/or decay of the norm of the wave function. Since the Hamiltonian we are considering is Hermitian in the absence of the imaginary vector potential, the norm of the wave function $\vec{\psi_0}$ remains one during the transportation. Therefore, the effect of the Aharonov-Bohm factor $e^{i \Phi}$ can be obtained directly by looking at the norm of the wave function in the final state.

To numerically demonstrate this non-Hermitian Aharonov-Bohm effect in real space, we consider the setup where we start from a localized wave packet around the center of the lattice, and then add external forces to make the wave packet move. As the trajectory of the wave packet forms a closed path, the change of the magnitude of the wave function is precisely related to the imaginary magnetic flux enclosed by the path. We now numerically demonstrate the effect.

\begin{figure}
    \centering
    \includegraphics[width=0.5\textwidth]{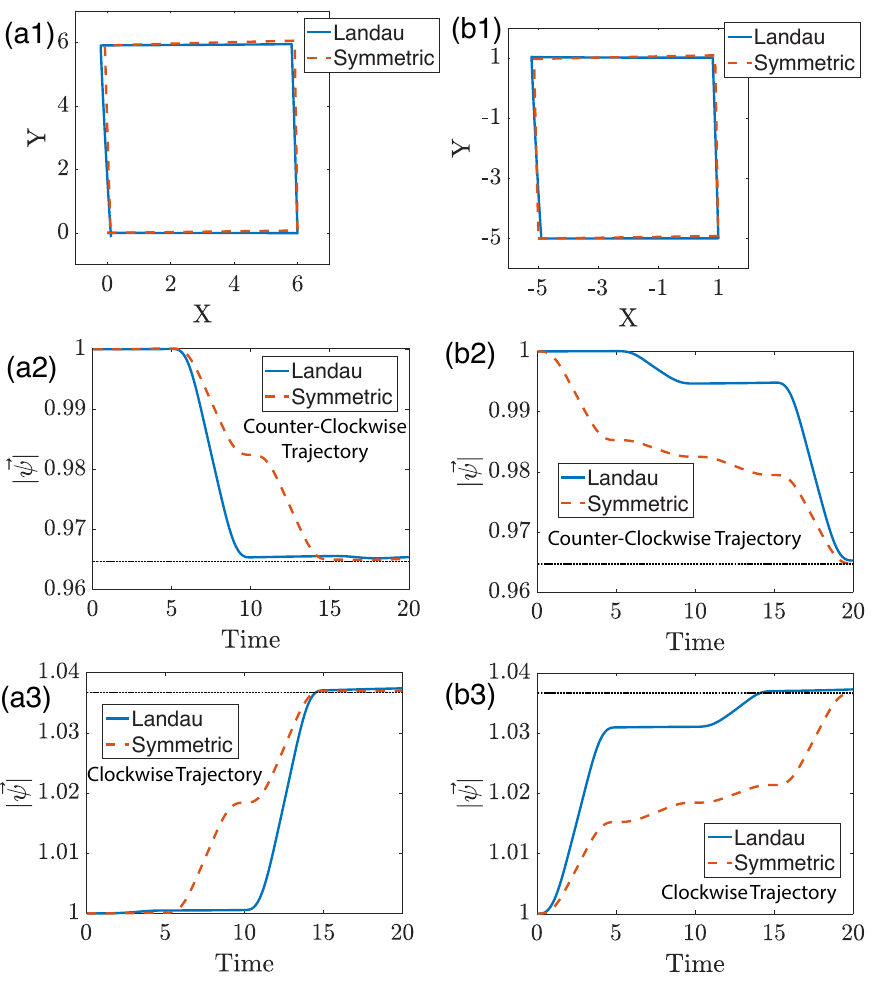}
    \caption{Simulation of the Aharonov-Bohm effect under imaginary magnetic fields for the lattice size of $50 \times 50$ with $B = 0.001i$ under open boundary conditions. (a1) The counter-clockwise trajectory of the mean position of the wave function, starting from the center of the lattice, which we take to be $(x_0, y_0)$ = (0,0). (a2) $|\vec{\psi}|$ as a function of time (in units of $1/J$) for the counter-clockwise trajectory. The horizontal dotted line is the theoretical value of the Aharonov-Bohm factor $e^{iB A_\mathrm{Area}} = e^{-\mathcal{B}A_\mathrm{Area}} \approx 0.965$. (a3) $|\vec{\psi}|$ as a function of time for the clockwise trajectory. The horizontal dotted line is at $e^{\mathcal{B}A_\mathrm{Area}} \approx 1.037$.
    We also performed the same simulation with a different starting point located at $(x_0, y_0) = (-5,-5)$, whose trajectory is plotted in (b1). The modulus of the wave function $|\vec{\psi}|$ as a function of time for the counter-clockwise trajectory and clockwise trajectory are plotted in (b2) and (b3), respectively, for the wave packet starting from $(x_0, y_0) = (-5,-5)$.
    We note that, although the $|\vec{\psi}|$ during the evolution is different for (a2), (a3) and (b2), (b3), the final values, which are solely determined by the enclosed imaginary magnetic flux, are common.
    In all plots, the solid lines and curves are for the Landau gauge, whereas the dashed lines and curves are for the symmetric gauge, respectively.}
    \label{fig:fourth}
\end{figure}

As an initial state, we choose a normalized Gaussian wave packet $\psi_{x,y} \propto e^{-\{(x-x_0)^2 + (y-y_0)^2)\}/(2\sigma^2)}$ centered around the point $(x_0, y_0)$ with the spread $\sigma = 5$.
We apply a force changing sinusoidally in time, created by a potential $V_{x,y} = E_x \sin (2\pi t/T) x + E_y \sin (2\pi t/T) y$,  so the wave packet makes a rectangular trajectory either in the counter-clockwise or clockwise direction. For the counter-clockwise trajectory, we apply
\begin{align}
    (E_x, E_y) =
    \begin{cases}
        (1,0) & \text{for }0 \le t \le T \\
        (0,1) & \text{for }T \le t \le 2T \\
        (-1,0) & \text{for }2T \le t \le 3T \\
        (0,-1) & \text{for }3T \le t \le 4T. \\
    \end{cases}
\end{align}
For the clockwise trajectory, we make the wave packet move in the opposite direction by applying
\begin{align}
    (E_x, E_y) =
    \begin{cases}
        (0,1) & \text{for }0 \le t \le T \\
        (1,0) & \text{for }T \le t \le 2T \\
        (0,-1) & \text{for }2T \le t \le 3T \\
        (-1,0) & \text{for }3T \le t \le 4T. \\
    \end{cases}
\end{align}
We use $T = 5/J$ in the numerical simulation. 

In the following numerical simulation, we choose a lattice of size $N_x = N_y = 50$ with an imaginary magnetic field $B = 0.001i$ under open boundary conditions. We use coordinates so $(x, y) = (0,0)$ is located at the center of the lattice. Starting from a wave packet centered around $(x_0, y_0) = (0,0)$, and evolving in time until $t = 4T$, the center of the wave packet forms rectangles as plotted in Fig.~\ref{fig:fourth}(a1) for the counter-clockwise trajectory. In Fig.~\ref{fig:fourth}(a2), we plot the modulus of the wave function, $|\vec{\psi}| \equiv \sqrt{\sum_{x,y}|\psi_{x,y}|^2}$, as a function of time for both the Landau and symmetric gauges for the counter-clockwise trajectory. The same quantity for the clockwise trajectory is also plotted in Fig.~\ref{fig:fourth}(a3). We see that during the time evolution, $|\vec{\psi}|$ is generally different between the two gauges, but it is the same after the closed trajectory is formed. 
The Aharonov-Bohm factor corresponding to this trajectory is $e^{i\Phi} = e^{iB A_\mathrm{Area}} = e^{-\mathcal{B}A_\mathrm{Area}} \approx 0.965$ for the counter-clockwise trajectory, where the area enclosed by the trajectory is $A_\mathrm{Area} \approx 36$ for both gauges.
For the clockwise trajectory, $e^{i\Phi} = e^{-iB A_\mathrm{Area}} = e^{\mathcal{B}A_\mathrm{Area}} \approx 1.037$.
Those values are also plotted in Figs.~\ref{fig:fourth}(a2) and~\ref{fig:fourth}(a3).
Although, both for counter-clockwise and clockwise trajectories, the time evolution of $|\vec{\psi}|$ depends on the choice of gauge, the final values of Figs.~\ref{fig:fourth}(a2) and~\ref{fig:fourth}(a3) agree well with the Aharonov-Bohm factor and are independent of the gauge choice.

The detailed time-dependence of $|\vec{\psi}|$ can also be understood from the Aharonov-Bohm factor during the evolution, which is $\exp \left( i \int_{\mathbf{r}_i, \mathcal{C}}^{\mathbf{r}_f} \mathbf{A}(\mathbf{r}^\prime)\cdot d\mathbf{r}^\prime\right)$. For example, during $0 \le t  \le T$, $|\vec{\psi}|$ remains to be 1 both for the Landau and symmetric gauges in Fig.~\ref{fig:fourth}(a2). This is because, both for the Landau gauge $\mathbf{A} = (0, Bx)$ and the symmetric gauge $\mathbf{A} = (-By/2, Bx/2)$, the line integral along the horizontal direction is zero if the wave packet starts from $(x_0, y_0) = (0,0)$.

This observation also implies that the time evolution of $|\vec{\psi}|$ depends on the initial position of the wavepacket, whereas the final values of $|\vec{\psi}|$ are independent of the initial position and the gauge one chooses. This has been numerically demonstrated in Figs.~\ref{fig:fourth}(b1)-\ref{fig:fourth}(b3), where the initial wavepacket is centered around $(x_0, y_0) = (-5,-5)$. Comparing Fig.~\ref{fig:fourth}(a2) with Fig.~\ref{fig:fourth}(b2) for counter-clockwise trajectories, and Fig.~\ref{fig:fourth}(a3) with Fig.~\ref{fig:fourth}(b3) for clockwise trajectories, we confirm that the final values of $|\vec{\psi}|$ are $e^{\pm i\Phi}$, whereas the values during the evolution depend on the initial position of the wave packet and the gauge choice. 

\section{Conclusion}
\label{sec:concl}
We have studied spectral and geometrical properties of two-dimensional lattices under a uniform imaginary magnetic field. Our results unveil features of imaginary magnetic fields which are intrinsically different from real magnetic fields, such as the impossibility to take periodic boundary conditions in both directions and non-convergence of the energy spectrum in the limit when both sides are taken large. On the other hand, there also are similarities to the real magnetic field, such as description in terms of the Harper equation and the analog of the Aharonov-Bohm effect. Although we focused on the cases of purely imaginary magnetic fields, general results presented in the paper, such as the non-Bloch band theory when increasing the length of one direction and the non-Hermitian Aharonov-Bohm effect, should be valid also for more general complex magnetic fields including both real and imaginary components.

Experimental realization of the imaginary magnetic field we discussed in the paper requires constructing an extended lattice with desired non-reciprocal hopping amplitudes, which can be achieved in the platforms of mechanical metamaterials~\cite{Brandenbourger:2019NatComm,Ghatak:2020PNAS} and electrical circuits~\cite{Helbig:2020NatPhys,Hofmann:2020PRR,Anandwade:2023PRA} in a straightforward manner.
The model, in the Landau gauge, can also be realized by aligning a collection of one-dimensional Hatano-Nelson models and coupling them; such a collection of Hatano-Nelson models can be realized through Floquet methods, e.g., through the method described in Ref.~\cite{Faugno:2022PRL}, in various photonic systems such as coupled waveguides or coupled resonators.
In these classical setups, change of the norm of the wavefunction due to the non-Hermitian Aharonov-Bohm effect is observable through the amplification and/or damping of corresponding signals.
There is also a recent theoretical proposal of an Aharonov-Bohm-type setting where the complex vector potential emerges after post selection~\cite{Paiva:2023NJP}. 

Our results provide a starting point toward the research field of non-Hermitian magnetic fields.
We have focused on physics on lattices; understanding unique properties under an imaginary magnetic field in continuous two-dimensional systems is also an open field of study. Moreover, understanding properties under more general gauge fields such as complex electromagnetic fields and non-Abelian gauge fields (e.g., spin-orbit coupling) is left for future study.

\begin{acknowledgments}
The authors thank Shuichi Murakami for helpful discussions on the non-Bloch band theory.
This work is supported by JSPS KAKENHI Grant No. JP20H01845, Grant No. JP21H01007, Grant No. JP21H01084, JST PRESTO Grant No. JPMJPR2353, and JST CREST Grant No.JPMJCR19T1.
\end{acknowledgments}

\appendix*
\section{Spectral bounds of matrices}
\label{sec:appA}
We show that the real part of the eigenvalues of a diagonalizable matrix $M$ is bounded by the eigenvalues of its symmetric component, and the imaginary part of the eigenvalues is bounded by the eigenvalues of its anti-symmetric components. For the discussion in the main text, we only need to consider the case where $M$ is a real diagonalizable matrix, but here we provide a more general property which holds even when $M$ is a complex diagonalizable matrix.

Assume that $M$ is a complex diagonalizable matrix. We then write
\begin{align}
    M = \frac{M + M^\dagger}{2} + \frac{M - M^\dagger}{2},
\end{align}
where $M^\dagger$ is the Hermitian conjugate of $M$. Defining the first term as $S = \frac{M + M^\dagger}{2}$ and the second term as $\frac{M - M^\dagger}{2}$, we can see that $S$ is a Hermitian matrix and $A$ is an anti-Hermitian matrix, namely, $S^\dagger = S$ and $A^\dagger = -A$.
We note that eigenvalues of a Hermitian matrix and an anti-Hermitian matrix are real and pure imaginary, respectively.

For any normalized vector $|v\rangle$, we see the following property:
\begin{align}
    \langle v | S | v \rangle &= \langle v| \frac{M + M^\dagger}{2} | v \rangle
    =
    \frac{1}{2}\left( \langle v | M | v\rangle + \langle v | M | v\rangle^* \right) \notag \\
    &=
    \mathrm{Re}\left( \langle v | M | v\rangle \right).
\end{align}
Since $S$ is a Hermitian matrix, the minimum value of the left hand side is given when $|v\rangle$ is an eigenvector of the minimum eigenvalue $\lambda_{\mathrm{min}}$ of $S$:
\begin{align}
    \mathrm{min} \langle v | S | v \rangle = \lambda_\mathrm{min}.
\end{align}
Similarly, the maximum value of the left-hand side is equal to the maximum eigenvalue $\lambda_{\mathrm{max}}$ of $S$.
Thus we obtain the following inequality:
\begin{align}
    \lambda_\mathrm{min} \le 
    \mathrm{Re}\left( \langle v | M | v\rangle \right)
    \le \lambda_\mathrm{max}.
\end{align}
Now, taking $|v\rangle$ to be an eigenvector of $M$ with an eigenvalue $E$, $M|v\rangle = E|v\rangle$, 
we finally obtain the desired inequality:
\begin{align}
    \lambda_\mathrm{min} \le \mathrm{Re}(E) \le \lambda_\mathrm{max}.
\end{align}
We note that the equality is not necessarily achieved because eigenvectors of $M$ are not orthogonal and there can be ways to minimize $\mathrm{Re}\left( \langle v | M | v\rangle \right)$ by choosing $|v\rangle$ to not be an eigenvector of $M$.

We now prove a similar relation for the spread of the eigenvalues of $M$ along the imaginary axis. 
The property of the imaginary part of the eigenvalues that we want to show can be mapped to the problem of the real part by writing
\begin{align}
    -iM = -iS -iA;
\end{align}
the second term $-iA$ is Hermitian and the first term $-iS$ is anti-Hermitian.
This implies that the spread of the eigenvalues of $-iM$ along the real axis is bounded by the real eigenvalues of $-iA$. Therefore, the spread of the eigenvalues of $M$ along the imaginary axis is bounded by the eigenvalues of $A$.

We have thus proved that, writing an arbitrary diagonalizable complex matrix $M$ in terms of its Hermitian and anti-Hermitian parts by $M = S + A$, the spectral range of $M$ along the real part is bounded by the eigenvalues of $S$ and the imaginary part is bounded by the eigenvalues of $A$.
When $M$ is a real matrix, the same result holds when writing $M = S + A$, with $S$ being a symmetric matrix and $A$ being an anti-symmetric matrix.

\bibliography{bibliography.bib}

\begin{thebibliography}{39}%
\makeatletter
\providecommand \@ifxundefined [1]{%
 \@ifx{#1\undefined}
}%
\providecommand \@ifnum [1]{%
 \ifnum #1\expandafter \@firstoftwo
 \else \expandafter \@secondoftwo
 \fi
}%
\providecommand \@ifx [1]{%
 \ifx #1\expandafter \@firstoftwo
 \else \expandafter \@secondoftwo
 \fi
}%
\providecommand \natexlab [1]{#1}%
\providecommand \enquote  [1]{``#1''}%
\providecommand \bibnamefont  [1]{#1}%
\providecommand \bibfnamefont [1]{#1}%
\providecommand \citenamefont [1]{#1}%
\providecommand \href@noop [0]{\@secondoftwo}%
\providecommand \href [0]{\begingroup \@sanitize@url \@href}%
\providecommand \@href[1]{\@@startlink{#1}\@@href}%
\providecommand \@@href[1]{\endgroup#1\@@endlink}%
\providecommand \@sanitize@url [0]{\catcode `\\12\catcode `\$12\catcode
  `\&12\catcode `\#12\catcode `\^12\catcode `\_12\catcode `\%12\relax}%
\providecommand \@@startlink[1]{}%
\providecommand \@@endlink[0]{}%
\providecommand \url  [0]{\begingroup\@sanitize@url \@url }%
\providecommand \@url [1]{\endgroup\@href {#1}{\urlprefix }}%
\providecommand \urlprefix  [0]{URL }%
\providecommand \Eprint [0]{\href }%
\providecommand \doibase [0]{https://doi.org/}%
\providecommand \selectlanguage [0]{\@gobble}%
\providecommand \bibinfo  [0]{\@secondoftwo}%
\providecommand \bibfield  [0]{\@secondoftwo}%
\providecommand \translation [1]{[#1]}%
\providecommand \BibitemOpen [0]{}%
\providecommand \bibitemStop [0]{}%
\providecommand \bibitemNoStop [0]{.\EOS\space}%
\providecommand \EOS [0]{\spacefactor3000\relax}%
\providecommand \BibitemShut  [1]{\csname bibitem#1\endcsname}%
\let\auto@bib@innerbib\@empty
\bibitem [{\citenamefont {Ashcroft}\ and\ \citenamefont
  {Mermin}(1976)}]{AshcroftMermin}%
  \BibitemOpen
  \bibfield  {author} {\bibinfo {author} {\bibfnamefont {N.~W.}\ \bibnamefont
  {Ashcroft}}\ and\ \bibinfo {author} {\bibfnamefont {N.~D.}\ \bibnamefont
  {Mermin}},\ }\href@noop {} {\emph {\bibinfo {title} {{S}olid {S}tate
  {P}hysics}}}\ (\bibinfo  {publisher} {Saunders College Publishing, New
  York},\ \bibinfo {year} {1976})\BibitemShut {NoStop}%
\bibitem [{\citenamefont {Klitzing}\ \emph {et~al.}(1980)\citenamefont
  {Klitzing}, \citenamefont {Dorda},\ and\ \citenamefont
  {Pepper}}]{Klitzing:1980PRL}%
  \BibitemOpen
  \bibfield  {author} {\bibinfo {author} {\bibfnamefont {K.~v.}\ \bibnamefont
  {Klitzing}}, \bibinfo {author} {\bibfnamefont {G.}~\bibnamefont {Dorda}},\
  and\ \bibinfo {author} {\bibfnamefont {M.}~\bibnamefont {Pepper}},\
  }\bibfield  {title} {\bibinfo {title} {New method for high-accuracy
  determination of the fine-structure constant based on quantized {H}all
  resistance},\ }\href {https://doi.org/10.1103/PhysRevLett.45.494} {\bibfield
  {journal} {\bibinfo  {journal} {Phys. Rev. Lett.}\ }\textbf {\bibinfo
  {volume} {45}},\ \bibinfo {pages} {494} (\bibinfo {year} {1980})}\BibitemShut
  {NoStop}%
\bibitem [{\citenamefont {Tsui}\ \emph {et~al.}(1982)\citenamefont {Tsui},
  \citenamefont {Stormer},\ and\ \citenamefont {Gossard}}]{Tsui:1982PRL}%
  \BibitemOpen
  \bibfield  {author} {\bibinfo {author} {\bibfnamefont {D.~C.}\ \bibnamefont
  {Tsui}}, \bibinfo {author} {\bibfnamefont {H.~L.}\ \bibnamefont {Stormer}},\
  and\ \bibinfo {author} {\bibfnamefont {A.~C.}\ \bibnamefont {Gossard}},\
  }\bibfield  {title} {\bibinfo {title} {Two-dimensional magnetotransport in
  the extreme quantum limit},\ }\href
  {https://doi.org/10.1103/PhysRevLett.48.1559} {\bibfield  {journal} {\bibinfo
   {journal} {Phys. Rev. Lett.}\ }\textbf {\bibinfo {volume} {48}},\ \bibinfo
  {pages} {1559} (\bibinfo {year} {1982})}\BibitemShut {NoStop}%
\bibitem [{\citenamefont {Harper}(1955)}]{Harper:1955PPS}%
  \BibitemOpen
  \bibfield  {author} {\bibinfo {author} {\bibfnamefont {P.~G.}\ \bibnamefont
  {Harper}},\ }\bibfield  {title} {\bibinfo {title} {Single band motion of
  conduction electrons in a uniform magnetic field},\ }\href
  {https://iopscience.iop.org/article/10.1088/0370-1298/68/10/304} {\bibfield
  {journal} {\bibinfo  {journal} {Proceedings of the Physical Society. Section
  A}\ }\textbf {\bibinfo {volume} {68}},\ \bibinfo {pages} {874} (\bibinfo
  {year} {1955})}\BibitemShut {NoStop}%
\bibitem [{\citenamefont {Hofstadter}(1976)}]{Hofstadter:1976PRB}%
  \BibitemOpen
  \bibfield  {author} {\bibinfo {author} {\bibfnamefont {D.~R.}\ \bibnamefont
  {Hofstadter}},\ }\bibfield  {title} {\bibinfo {title} {Energy levels and wave
  functions of bloch electrons in rational and irrational magnetic fields},\
  }\href {https://doi.org/10.1103/PhysRevB.14.2239} {\bibfield  {journal}
  {\bibinfo  {journal} {Phys. Rev. B}\ }\textbf {\bibinfo {volume} {14}},\
  \bibinfo {pages} {2239} (\bibinfo {year} {1976})}\BibitemShut {NoStop}%
\bibitem [{\citenamefont {Thouless}\ \emph {et~al.}(1982)\citenamefont
  {Thouless}, \citenamefont {Kohmoto}, \citenamefont {Nightingale},\ and\
  \citenamefont {den Nijs}}]{Thouless:1982PRL}%
  \BibitemOpen
  \bibfield  {author} {\bibinfo {author} {\bibfnamefont {D.~J.}\ \bibnamefont
  {Thouless}}, \bibinfo {author} {\bibfnamefont {M.}~\bibnamefont {Kohmoto}},
  \bibinfo {author} {\bibfnamefont {M.~P.}\ \bibnamefont {Nightingale}},\ and\
  \bibinfo {author} {\bibfnamefont {M.}~\bibnamefont {den Nijs}},\ }\bibfield
  {title} {\bibinfo {title} {Quantized {H}all conductance in a two-dimensional
  periodic potential},\ }\href {https://doi.org/10.1103/PhysRevLett.49.405}
  {\bibfield  {journal} {\bibinfo  {journal} {Phys. Rev. Lett.}\ }\textbf
  {\bibinfo {volume} {49}},\ \bibinfo {pages} {405} (\bibinfo {year}
  {1982})}\BibitemShut {NoStop}%
\bibitem [{\citenamefont {Ashida}\ \emph {et~al.}(2020)\citenamefont {Ashida},
  \citenamefont {Gong},\ and\ \citenamefont {Ueda}}]{Ashida:2020AdvPhys}%
  \BibitemOpen
  \bibfield  {author} {\bibinfo {author} {\bibfnamefont {Y.}~\bibnamefont
  {Ashida}}, \bibinfo {author} {\bibfnamefont {Z.}~\bibnamefont {Gong}},\ and\
  \bibinfo {author} {\bibfnamefont {M.}~\bibnamefont {Ueda}},\ }\bibfield
  {title} {\bibinfo {title} {Non-{H}ermitian physics},\ }\href
  {https://www.tandfonline.com/doi/abs/10.1080/00018732.2021.1876991?journalCode=tadp20}
  {\bibfield  {journal} {\bibinfo  {journal} {Advances in Physics}\ }\textbf
  {\bibinfo {volume} {69}},\ \bibinfo {pages} {249} (\bibinfo {year}
  {2020})}\BibitemShut {NoStop}%
\bibitem [{\citenamefont {Bergholtz}\ \emph {et~al.}(2021)\citenamefont
  {Bergholtz}, \citenamefont {Budich},\ and\ \citenamefont
  {Kunst}}]{Bergholtz:2021RMP}%
  \BibitemOpen
  \bibfield  {author} {\bibinfo {author} {\bibfnamefont {E.~J.}\ \bibnamefont
  {Bergholtz}}, \bibinfo {author} {\bibfnamefont {J.~C.}\ \bibnamefont
  {Budich}},\ and\ \bibinfo {author} {\bibfnamefont {F.~K.}\ \bibnamefont
  {Kunst}},\ }\bibfield  {title} {\bibinfo {title} {Exceptional topology of
  non-{H}ermitian systems},\ }\href
  {https://doi.org/10.1103/RevModPhys.93.015005} {\bibfield  {journal}
  {\bibinfo  {journal} {Rev. Mod. Phys.}\ }\textbf {\bibinfo {volume} {93}},\
  \bibinfo {pages} {015005} (\bibinfo {year} {2021})}\BibitemShut {NoStop}%
\bibitem [{\citenamefont {Hatano}\ and\ \citenamefont
  {Nelson}(1996)}]{Hatano:1996PRL}%
  \BibitemOpen
  \bibfield  {author} {\bibinfo {author} {\bibfnamefont {N.}~\bibnamefont
  {Hatano}}\ and\ \bibinfo {author} {\bibfnamefont {D.~R.}\ \bibnamefont
  {Nelson}},\ }\bibfield  {title} {\bibinfo {title} {Localization transitions
  in non-{H}ermitian quantum mechanics},\ }\href
  {https://doi.org/10.1103/PhysRevLett.77.570} {\bibfield  {journal} {\bibinfo
  {journal} {Phys. Rev. Lett.}\ }\textbf {\bibinfo {volume} {77}},\ \bibinfo
  {pages} {570} (\bibinfo {year} {1996})}\BibitemShut {NoStop}%
\bibitem [{\citenamefont {Hatano}\ and\ \citenamefont
  {Nelson}(1997)}]{Hatano:1997PRB}%
  \BibitemOpen
  \bibfield  {author} {\bibinfo {author} {\bibfnamefont {N.}~\bibnamefont
  {Hatano}}\ and\ \bibinfo {author} {\bibfnamefont {D.~R.}\ \bibnamefont
  {Nelson}},\ }\bibfield  {title} {\bibinfo {title} {Vortex pinning and
  non-{H}ermitian quantum mechanics},\ }\href
  {https://doi.org/10.1103/PhysRevB.56.8651} {\bibfield  {journal} {\bibinfo
  {journal} {Phys. Rev. B}\ }\textbf {\bibinfo {volume} {56}},\ \bibinfo
  {pages} {8651} (\bibinfo {year} {1997})}\BibitemShut {NoStop}%
\bibitem [{\citenamefont {Gong}\ \emph {et~al.}(2018)\citenamefont {Gong},
  \citenamefont {Ashida}, \citenamefont {Kawabata}, \citenamefont {Takasan},
  \citenamefont {Higashikawa},\ and\ \citenamefont {Ueda}}]{Gong:2018PRX}%
  \BibitemOpen
  \bibfield  {author} {\bibinfo {author} {\bibfnamefont {Z.}~\bibnamefont
  {Gong}}, \bibinfo {author} {\bibfnamefont {Y.}~\bibnamefont {Ashida}},
  \bibinfo {author} {\bibfnamefont {K.}~\bibnamefont {Kawabata}}, \bibinfo
  {author} {\bibfnamefont {K.}~\bibnamefont {Takasan}}, \bibinfo {author}
  {\bibfnamefont {S.}~\bibnamefont {Higashikawa}},\ and\ \bibinfo {author}
  {\bibfnamefont {M.}~\bibnamefont {Ueda}},\ }\bibfield  {title} {\bibinfo
  {title} {Topological phases of non-{H}ermitian systems},\ }\href
  {https://doi.org/10.1103/PhysRevX.8.031079} {\bibfield  {journal} {\bibinfo
  {journal} {Phys. Rev. X}\ }\textbf {\bibinfo {volume} {8}},\ \bibinfo {pages}
  {031079} (\bibinfo {year} {2018})}\BibitemShut {NoStop}%
\bibitem [{\citenamefont {Kawabata}\ \emph {et~al.}(2019)\citenamefont
  {Kawabata}, \citenamefont {Shiozaki}, \citenamefont {Ueda},\ and\
  \citenamefont {Sato}}]{Kawabata:2019PRX}%
  \BibitemOpen
  \bibfield  {author} {\bibinfo {author} {\bibfnamefont {K.}~\bibnamefont
  {Kawabata}}, \bibinfo {author} {\bibfnamefont {K.}~\bibnamefont {Shiozaki}},
  \bibinfo {author} {\bibfnamefont {M.}~\bibnamefont {Ueda}},\ and\ \bibinfo
  {author} {\bibfnamefont {M.}~\bibnamefont {Sato}},\ }\bibfield  {title}
  {\bibinfo {title} {Symmetry and topology in non-{H}ermitian physics},\ }\href
  {https://doi.org/10.1103/PhysRevX.9.041015} {\bibfield  {journal} {\bibinfo
  {journal} {Phys. Rev. X}\ }\textbf {\bibinfo {volume} {9}},\ \bibinfo {pages}
  {041015} (\bibinfo {year} {2019})}\BibitemShut {NoStop}%
\bibitem [{\citenamefont {Borgnia}\ \emph {et~al.}(2020)\citenamefont
  {Borgnia}, \citenamefont {Kruchkov},\ and\ \citenamefont
  {Slager}}]{Borgnia:2020PRL}%
  \BibitemOpen
  \bibfield  {author} {\bibinfo {author} {\bibfnamefont {D.~S.}\ \bibnamefont
  {Borgnia}}, \bibinfo {author} {\bibfnamefont {A.~J.}\ \bibnamefont
  {Kruchkov}},\ and\ \bibinfo {author} {\bibfnamefont {R.-J.}\ \bibnamefont
  {Slager}},\ }\bibfield  {title} {\bibinfo {title} {Non-{H}ermitian boundary
  modes and topology},\ }\href {https://doi.org/10.1103/PhysRevLett.124.056802}
  {\bibfield  {journal} {\bibinfo  {journal} {Phys. Rev. Lett.}\ }\textbf
  {\bibinfo {volume} {124}},\ \bibinfo {pages} {056802} (\bibinfo {year}
  {2020})}\BibitemShut {NoStop}%
\bibitem [{\citenamefont {Okuma}\ \emph {et~al.}(2020)\citenamefont {Okuma},
  \citenamefont {Kawabata}, \citenamefont {Shiozaki},\ and\ \citenamefont
  {Sato}}]{Okuma:2020PRL}%
  \BibitemOpen
  \bibfield  {author} {\bibinfo {author} {\bibfnamefont {N.}~\bibnamefont
  {Okuma}}, \bibinfo {author} {\bibfnamefont {K.}~\bibnamefont {Kawabata}},
  \bibinfo {author} {\bibfnamefont {K.}~\bibnamefont {Shiozaki}},\ and\
  \bibinfo {author} {\bibfnamefont {M.}~\bibnamefont {Sato}},\ }\bibfield
  {title} {\bibinfo {title} {Topological origin of non-{H}ermitian skin
  effects},\ }\href {https://doi.org/10.1103/PhysRevLett.124.086801} {\bibfield
   {journal} {\bibinfo  {journal} {Phys. Rev. Lett.}\ }\textbf {\bibinfo
  {volume} {124}},\ \bibinfo {pages} {086801} (\bibinfo {year}
  {2020})}\BibitemShut {NoStop}%
\bibitem [{\citenamefont {Lin}\ \emph {et~al.}(2023)\citenamefont {Lin},
  \citenamefont {Tai}, \citenamefont {Li},\ and\ \citenamefont
  {Lee}}]{Lin:2023FrontPhys}%
  \BibitemOpen
  \bibfield  {author} {\bibinfo {author} {\bibfnamefont {R.}~\bibnamefont
  {Lin}}, \bibinfo {author} {\bibfnamefont {T.}~\bibnamefont {Tai}}, \bibinfo
  {author} {\bibfnamefont {L.}~\bibnamefont {Li}},\ and\ \bibinfo {author}
  {\bibfnamefont {C.~H.}\ \bibnamefont {Lee}},\ }\bibfield  {title} {\bibinfo
  {title} {Topological non-{H}ermitian skin effect},\ }\href
  {https://link.springer.com/article/10.1007/s11467-023-1309-z} {\bibfield
  {journal} {\bibinfo  {journal} {Frontiers of Physics}\ }\textbf {\bibinfo
  {volume} {18}},\ \bibinfo {pages} {53605} (\bibinfo {year}
  {2023})}\BibitemShut {NoStop}%
\bibitem [{\citenamefont {Brandenbourger}\ \emph {et~al.}(2019)\citenamefont
  {Brandenbourger}, \citenamefont {Locsin}, \citenamefont {Lerner},\ and\
  \citenamefont {Coulais}}]{Brandenbourger:2019NatComm}%
  \BibitemOpen
  \bibfield  {author} {\bibinfo {author} {\bibfnamefont {M.}~\bibnamefont
  {Brandenbourger}}, \bibinfo {author} {\bibfnamefont {X.}~\bibnamefont
  {Locsin}}, \bibinfo {author} {\bibfnamefont {E.}~\bibnamefont {Lerner}},\
  and\ \bibinfo {author} {\bibfnamefont {C.}~\bibnamefont {Coulais}},\
  }\bibfield  {title} {\bibinfo {title} {Non-reciprocal robotic
  metamaterials},\ }\href {https://www.nature.com/articles/s41467-019-12599-3}
  {\bibfield  {journal} {\bibinfo  {journal} {Nature communications}\ }\textbf
  {\bibinfo {volume} {10}},\ \bibinfo {pages} {4608} (\bibinfo {year}
  {2019})}\BibitemShut {NoStop}%
\bibitem [{\citenamefont {Ghatak}\ \emph {et~al.}(2020)\citenamefont {Ghatak},
  \citenamefont {Brandenbourger}, \citenamefont {Van~Wezel},\ and\
  \citenamefont {Coulais}}]{Ghatak:2020PNAS}%
  \BibitemOpen
  \bibfield  {author} {\bibinfo {author} {\bibfnamefont {A.}~\bibnamefont
  {Ghatak}}, \bibinfo {author} {\bibfnamefont {M.}~\bibnamefont
  {Brandenbourger}}, \bibinfo {author} {\bibfnamefont {J.}~\bibnamefont
  {Van~Wezel}},\ and\ \bibinfo {author} {\bibfnamefont {C.}~\bibnamefont
  {Coulais}},\ }\bibfield  {title} {\bibinfo {title} {Observation of
  non-{H}ermitian topology and its bulk--edge correspondence in an active
  mechanical metamaterial},\ }\href
  {https://www.pnas.org/doi/full/10.1073/pnas.2010580117} {\bibfield  {journal}
  {\bibinfo  {journal} {Proceedings of the National Academy of Sciences}\
  }\textbf {\bibinfo {volume} {117}},\ \bibinfo {pages} {29561} (\bibinfo
  {year} {2020})}\BibitemShut {NoStop}%
\bibitem [{\citenamefont {Helbig}\ \emph {et~al.}(2020)\citenamefont {Helbig},
  \citenamefont {Hofmann}, \citenamefont {Imhof}, \citenamefont {Abdelghany},
  \citenamefont {Kiessling}, \citenamefont {Molenkamp}, \citenamefont {Lee},
  \citenamefont {Szameit}, \citenamefont {Greiter},\ and\ \citenamefont
  {Thomale}}]{Helbig:2020NatPhys}%
  \BibitemOpen
  \bibfield  {author} {\bibinfo {author} {\bibfnamefont {T.}~\bibnamefont
  {Helbig}}, \bibinfo {author} {\bibfnamefont {T.}~\bibnamefont {Hofmann}},
  \bibinfo {author} {\bibfnamefont {S.}~\bibnamefont {Imhof}}, \bibinfo
  {author} {\bibfnamefont {M.}~\bibnamefont {Abdelghany}}, \bibinfo {author}
  {\bibfnamefont {T.}~\bibnamefont {Kiessling}}, \bibinfo {author}
  {\bibfnamefont {L.}~\bibnamefont {Molenkamp}}, \bibinfo {author}
  {\bibfnamefont {C.}~\bibnamefont {Lee}}, \bibinfo {author} {\bibfnamefont
  {A.}~\bibnamefont {Szameit}}, \bibinfo {author} {\bibfnamefont
  {M.}~\bibnamefont {Greiter}},\ and\ \bibinfo {author} {\bibfnamefont
  {R.}~\bibnamefont {Thomale}},\ }\bibfield  {title} {\bibinfo {title}
  {Generalized bulk--boundary correspondence in non-{H}ermitian topolectrical
  circuits},\ }\href {https://www.nature.com/articles/s41567-020-0922-9}
  {\bibfield  {journal} {\bibinfo  {journal} {Nature Physics}\ }\textbf
  {\bibinfo {volume} {16}},\ \bibinfo {pages} {747} (\bibinfo {year}
  {2020})}\BibitemShut {NoStop}%
\bibitem [{\citenamefont {Hofmann}\ \emph {et~al.}(2020)\citenamefont
  {Hofmann}, \citenamefont {Helbig}, \citenamefont {Schindler}, \citenamefont
  {Salgo}, \citenamefont {Brzezi\ifmmode~\acute{n}\else \'{n}\fi{}ska},
  \citenamefont {Greiter}, \citenamefont {Kiessling}, \citenamefont {Wolf},
  \citenamefont {Vollhardt}, \citenamefont {Kaba\ifmmode~\check{s}\else
  \v{s}\fi{}i}, \citenamefont {Lee}, \citenamefont {Bilu\ifmmode \check{s}\else
  \v{s}\fi{}i\ifmmode~\acute{c}\else \'{c}\fi{}}, \citenamefont {Thomale},\
  and\ \citenamefont {Neupert}}]{Hofmann:2020PRR}%
  \BibitemOpen
  \bibfield  {author} {\bibinfo {author} {\bibfnamefont {T.}~\bibnamefont
  {Hofmann}}, \bibinfo {author} {\bibfnamefont {T.}~\bibnamefont {Helbig}},
  \bibinfo {author} {\bibfnamefont {F.}~\bibnamefont {Schindler}}, \bibinfo
  {author} {\bibfnamefont {N.}~\bibnamefont {Salgo}}, \bibinfo {author}
  {\bibfnamefont {M.}~\bibnamefont {Brzezi\ifmmode~\acute{n}\else
  \'{n}\fi{}ska}}, \bibinfo {author} {\bibfnamefont {M.}~\bibnamefont
  {Greiter}}, \bibinfo {author} {\bibfnamefont {T.}~\bibnamefont {Kiessling}},
  \bibinfo {author} {\bibfnamefont {D.}~\bibnamefont {Wolf}}, \bibinfo {author}
  {\bibfnamefont {A.}~\bibnamefont {Vollhardt}}, \bibinfo {author}
  {\bibfnamefont {A.}~\bibnamefont {Kaba\ifmmode~\check{s}\else \v{s}\fi{}i}},
  \bibinfo {author} {\bibfnamefont {C.~H.}\ \bibnamefont {Lee}}, \bibinfo
  {author} {\bibfnamefont {A.}~\bibnamefont {Bilu\ifmmode \check{s}\else
  \v{s}\fi{}i\ifmmode~\acute{c}\else \'{c}\fi{}}}, \bibinfo {author}
  {\bibfnamefont {R.}~\bibnamefont {Thomale}},\ and\ \bibinfo {author}
  {\bibfnamefont {T.}~\bibnamefont {Neupert}},\ }\bibfield  {title} {\bibinfo
  {title} {Reciprocal skin effect and its realization in a topolectrical
  circuit},\ }\href {https://doi.org/10.1103/PhysRevResearch.2.023265}
  {\bibfield  {journal} {\bibinfo  {journal} {Phys. Rev. Res.}\ }\textbf
  {\bibinfo {volume} {2}},\ \bibinfo {pages} {023265} (\bibinfo {year}
  {2020})}\BibitemShut {NoStop}%
\bibitem [{\citenamefont {Weidemann}\ \emph {et~al.}(2020)\citenamefont
  {Weidemann}, \citenamefont {Kremer}, \citenamefont {Helbig}, \citenamefont
  {Hofmann}, \citenamefont {Stegmaier}, \citenamefont {Greiter}, \citenamefont
  {Thomale},\ and\ \citenamefont {Szameit}}]{Weidemann:2020Science}%
  \BibitemOpen
  \bibfield  {author} {\bibinfo {author} {\bibfnamefont {S.}~\bibnamefont
  {Weidemann}}, \bibinfo {author} {\bibfnamefont {M.}~\bibnamefont {Kremer}},
  \bibinfo {author} {\bibfnamefont {T.}~\bibnamefont {Helbig}}, \bibinfo
  {author} {\bibfnamefont {T.}~\bibnamefont {Hofmann}}, \bibinfo {author}
  {\bibfnamefont {A.}~\bibnamefont {Stegmaier}}, \bibinfo {author}
  {\bibfnamefont {M.}~\bibnamefont {Greiter}}, \bibinfo {author} {\bibfnamefont
  {R.}~\bibnamefont {Thomale}},\ and\ \bibinfo {author} {\bibfnamefont
  {A.}~\bibnamefont {Szameit}},\ }\bibfield  {title} {\bibinfo {title}
  {Topological funneling of light},\ }\href
  {https://www.science.org/doi/10.1126/science.aaz8727} {\bibfield  {journal}
  {\bibinfo  {journal} {Science}\ }\textbf {\bibinfo {volume} {368}},\ \bibinfo
  {pages} {311} (\bibinfo {year} {2020})}\BibitemShut {NoStop}%
\bibitem [{\citenamefont {Xiao}\ \emph {et~al.}(2020)\citenamefont {Xiao},
  \citenamefont {Deng}, \citenamefont {Wang}, \citenamefont {Zhu},
  \citenamefont {Wang}, \citenamefont {Yi},\ and\ \citenamefont
  {Xue}}]{Xiao:2020NatPhys}%
  \BibitemOpen
  \bibfield  {author} {\bibinfo {author} {\bibfnamefont {L.}~\bibnamefont
  {Xiao}}, \bibinfo {author} {\bibfnamefont {T.}~\bibnamefont {Deng}}, \bibinfo
  {author} {\bibfnamefont {K.}~\bibnamefont {Wang}}, \bibinfo {author}
  {\bibfnamefont {G.}~\bibnamefont {Zhu}}, \bibinfo {author} {\bibfnamefont
  {Z.}~\bibnamefont {Wang}}, \bibinfo {author} {\bibfnamefont {W.}~\bibnamefont
  {Yi}},\ and\ \bibinfo {author} {\bibfnamefont {P.}~\bibnamefont {Xue}},\
  }\bibfield  {title} {\bibinfo {title} {Non-{H}ermitian bulk--boundary
  correspondence in quantum dynamics},\ }\href
  {https://www.nature.com/articles/s41567-020-0836-6} {\bibfield  {journal}
  {\bibinfo  {journal} {Nature Physics}\ }\textbf {\bibinfo {volume} {16}},\
  \bibinfo {pages} {761} (\bibinfo {year} {2020})}\BibitemShut {NoStop}%
\bibitem [{\citenamefont {Xiao}\ \emph {et~al.}(2021)\citenamefont {Xiao},
  \citenamefont {Deng}, \citenamefont {Wang}, \citenamefont {Wang},
  \citenamefont {Yi},\ and\ \citenamefont {Xue}}]{Xiao:2021PRL}%
  \BibitemOpen
  \bibfield  {author} {\bibinfo {author} {\bibfnamefont {L.}~\bibnamefont
  {Xiao}}, \bibinfo {author} {\bibfnamefont {T.}~\bibnamefont {Deng}}, \bibinfo
  {author} {\bibfnamefont {K.}~\bibnamefont {Wang}}, \bibinfo {author}
  {\bibfnamefont {Z.}~\bibnamefont {Wang}}, \bibinfo {author} {\bibfnamefont
  {W.}~\bibnamefont {Yi}},\ and\ \bibinfo {author} {\bibfnamefont
  {P.}~\bibnamefont {Xue}},\ }\bibfield  {title} {\bibinfo {title} {Observation
  of non-bloch parity-time symmetry and exceptional points},\ }\href
  {https://doi.org/10.1103/PhysRevLett.126.230402} {\bibfield  {journal}
  {\bibinfo  {journal} {Phys. Rev. Lett.}\ }\textbf {\bibinfo {volume} {126}},\
  \bibinfo {pages} {230402} (\bibinfo {year} {2021})}\BibitemShut {NoStop}%
\bibitem [{\citenamefont {Liang}\ \emph {et~al.}(2022)\citenamefont {Liang},
  \citenamefont {Xie}, \citenamefont {Dong}, \citenamefont {Li}, \citenamefont
  {Li}, \citenamefont {Gadway}, \citenamefont {Yi},\ and\ \citenamefont
  {Yan}}]{Liang:2022PRL}%
  \BibitemOpen
  \bibfield  {author} {\bibinfo {author} {\bibfnamefont {Q.}~\bibnamefont
  {Liang}}, \bibinfo {author} {\bibfnamefont {D.}~\bibnamefont {Xie}}, \bibinfo
  {author} {\bibfnamefont {Z.}~\bibnamefont {Dong}}, \bibinfo {author}
  {\bibfnamefont {H.}~\bibnamefont {Li}}, \bibinfo {author} {\bibfnamefont
  {H.}~\bibnamefont {Li}}, \bibinfo {author} {\bibfnamefont {B.}~\bibnamefont
  {Gadway}}, \bibinfo {author} {\bibfnamefont {W.}~\bibnamefont {Yi}},\ and\
  \bibinfo {author} {\bibfnamefont {B.}~\bibnamefont {Yan}},\ }\bibfield
  {title} {\bibinfo {title} {Dynamic signatures of non-{H}ermitian skin effect
  and topology in ultracold atoms},\ }\href
  {https://doi.org/10.1103/PhysRevLett.129.070401} {\bibfield  {journal}
  {\bibinfo  {journal} {Phys. Rev. Lett.}\ }\textbf {\bibinfo {volume} {129}},\
  \bibinfo {pages} {070401} (\bibinfo {year} {2022})}\BibitemShut {NoStop}%
\bibitem [{\citenamefont {Fukui}\ and\ \citenamefont
  {Kawakami}(1998)}]{PhysRevB.58.16051}%
  \BibitemOpen
  \bibfield  {author} {\bibinfo {author} {\bibfnamefont {T.}~\bibnamefont
  {Fukui}}\ and\ \bibinfo {author} {\bibfnamefont {N.}~\bibnamefont
  {Kawakami}},\ }\bibfield  {title} {\bibinfo {title} {Breakdown of the {M}ott
  insulator: Exact solution of an asymmetric {H}ubbard model},\ }\href
  {https://doi.org/10.1103/PhysRevB.58.16051} {\bibfield  {journal} {\bibinfo
  {journal} {Phys. Rev. B}\ }\textbf {\bibinfo {volume} {58}},\ \bibinfo
  {pages} {16051} (\bibinfo {year} {1998})}\BibitemShut {NoStop}%
\bibitem [{\citenamefont {Oka}\ and\ \citenamefont
  {Aoki}(2010)}]{PhysRevB.81.033103}%
  \BibitemOpen
  \bibfield  {author} {\bibinfo {author} {\bibfnamefont {T.}~\bibnamefont
  {Oka}}\ and\ \bibinfo {author} {\bibfnamefont {H.}~\bibnamefont {Aoki}},\
  }\bibfield  {title} {\bibinfo {title} {Dielectric breakdown in a {M}ott
  insulator: Many-body {S}chwinger-{L}andau-{Z}ener mechanism studied with a
  generalized {B}ethe ansatz},\ }\href
  {https://doi.org/10.1103/PhysRevB.81.033103} {\bibfield  {journal} {\bibinfo
  {journal} {Phys. Rev. B}\ }\textbf {\bibinfo {volume} {81}},\ \bibinfo
  {pages} {033103} (\bibinfo {year} {2010})}\BibitemShut {NoStop}%
\bibitem [{\citenamefont {Yamauchi}\ \emph {et~al.}(2020)\citenamefont
  {Yamauchi}, \citenamefont {Hayata}, \citenamefont {Uwamichi}, \citenamefont
  {Ozawa},\ and\ \citenamefont {Kawaguchi}}]{Yamauchi:2020arXiv}%
  \BibitemOpen
  \bibfield  {author} {\bibinfo {author} {\bibfnamefont {L.}~\bibnamefont
  {Yamauchi}}, \bibinfo {author} {\bibfnamefont {T.}~\bibnamefont {Hayata}},
  \bibinfo {author} {\bibfnamefont {M.}~\bibnamefont {Uwamichi}}, \bibinfo
  {author} {\bibfnamefont {T.}~\bibnamefont {Ozawa}},\ and\ \bibinfo {author}
  {\bibfnamefont {K.}~\bibnamefont {Kawaguchi}},\ }\bibfield  {title} {\bibinfo
  {title} {Chirality-driven edge flow and non-{H}ermitian topology in active
  nematic cells},\ }\href {https://arxiv.org/abs/2008.10852} {\bibfield
  {journal} {\bibinfo  {journal} {arXiv preprint arXiv:2008.10852}\ } (\bibinfo
  {year} {2020})}\BibitemShut {NoStop}%
\bibitem [{\citenamefont {Zou}\ \emph {et~al.}(2021)\citenamefont {Zou},
  \citenamefont {Chen}, \citenamefont {He}, \citenamefont {Bao}, \citenamefont
  {Lee}, \citenamefont {Sun},\ and\ \citenamefont {Zhang}}]{Zou:2021NatComm}%
  \BibitemOpen
  \bibfield  {author} {\bibinfo {author} {\bibfnamefont {D.}~\bibnamefont
  {Zou}}, \bibinfo {author} {\bibfnamefont {T.}~\bibnamefont {Chen}}, \bibinfo
  {author} {\bibfnamefont {W.}~\bibnamefont {He}}, \bibinfo {author}
  {\bibfnamefont {J.}~\bibnamefont {Bao}}, \bibinfo {author} {\bibfnamefont
  {C.~H.}\ \bibnamefont {Lee}}, \bibinfo {author} {\bibfnamefont
  {H.}~\bibnamefont {Sun}},\ and\ \bibinfo {author} {\bibfnamefont
  {X.}~\bibnamefont {Zhang}},\ }\bibfield  {title} {\bibinfo {title}
  {Observation of hybrid higher-order skin-topological effect in
  non-{H}ermitian topolectrical circuits},\ }\href@noop {} {\bibfield
  {journal} {\bibinfo  {journal} {Nature Communications}\ }\textbf {\bibinfo
  {volume} {12}},\ \bibinfo {pages} {7201} (\bibinfo {year}
  {2021})}\BibitemShut {NoStop}%
\bibitem [{\citenamefont {Palacios}\ \emph {et~al.}(2021)\citenamefont
  {Palacios}, \citenamefont {Tchoumakov}, \citenamefont {Guix}, \citenamefont
  {Pagonabarraga}, \citenamefont {S{\'a}nchez},\ and\ \citenamefont
  {G.~Grushin}}]{Palacios:2021NatComm}%
  \BibitemOpen
  \bibfield  {author} {\bibinfo {author} {\bibfnamefont {L.~S.}\ \bibnamefont
  {Palacios}}, \bibinfo {author} {\bibfnamefont {S.}~\bibnamefont
  {Tchoumakov}}, \bibinfo {author} {\bibfnamefont {M.}~\bibnamefont {Guix}},
  \bibinfo {author} {\bibfnamefont {I.}~\bibnamefont {Pagonabarraga}}, \bibinfo
  {author} {\bibfnamefont {S.}~\bibnamefont {S{\'a}nchez}},\ and\ \bibinfo
  {author} {\bibfnamefont {A.}~\bibnamefont {G.~Grushin}},\ }\bibfield  {title}
  {\bibinfo {title} {Guided accumulation of active particles by topological
  design of a second-order skin effect},\ }\href
  {https://www.nature.com/articles/s41467-021-24948-2} {\bibfield  {journal}
  {\bibinfo  {journal} {Nature Communications}\ }\textbf {\bibinfo {volume}
  {12}},\ \bibinfo {pages} {4691} (\bibinfo {year} {2021})}\BibitemShut
  {NoStop}%
\bibitem [{\citenamefont {Shang}\ \emph {et~al.}(2022)\citenamefont {Shang},
  \citenamefont {Liu}, \citenamefont {Shao}, \citenamefont {Han}, \citenamefont
  {Zang}, \citenamefont {Zhang}, \citenamefont {Salama}, \citenamefont {Gao},
  \citenamefont {Lee}, \citenamefont {Thomale} \emph
  {et~al.}}]{Shang:2022AdvSci}%
  \BibitemOpen
  \bibfield  {author} {\bibinfo {author} {\bibfnamefont {C.}~\bibnamefont
  {Shang}}, \bibinfo {author} {\bibfnamefont {S.}~\bibnamefont {Liu}}, \bibinfo
  {author} {\bibfnamefont {R.}~\bibnamefont {Shao}}, \bibinfo {author}
  {\bibfnamefont {P.}~\bibnamefont {Han}}, \bibinfo {author} {\bibfnamefont
  {X.}~\bibnamefont {Zang}}, \bibinfo {author} {\bibfnamefont {X.}~\bibnamefont
  {Zhang}}, \bibinfo {author} {\bibfnamefont {K.~N.}\ \bibnamefont {Salama}},
  \bibinfo {author} {\bibfnamefont {W.}~\bibnamefont {Gao}}, \bibinfo {author}
  {\bibfnamefont {C.~H.}\ \bibnamefont {Lee}}, \bibinfo {author} {\bibfnamefont
  {R.}~\bibnamefont {Thomale}}, \emph {et~al.},\ }\bibfield  {title} {\bibinfo
  {title} {Experimental identification of the second-order non-{H}ermitian skin
  effect with physics-graph-informed machine learning},\ }\href
  {https://onlinelibrary.wiley.com/doi/10.1002/advs.202202922} {\bibfield
  {journal} {\bibinfo  {journal} {Advanced Science}\ }\textbf {\bibinfo
  {volume} {9}},\ \bibinfo {pages} {2202922} (\bibinfo {year}
  {2022})}\BibitemShut {NoStop}%
\bibitem [{\citenamefont {Yao}\ and\ \citenamefont {Wang}(2018)}]{Yao:2018PRL}%
  \BibitemOpen
  \bibfield  {author} {\bibinfo {author} {\bibfnamefont {S.}~\bibnamefont
  {Yao}}\ and\ \bibinfo {author} {\bibfnamefont {Z.}~\bibnamefont {Wang}},\
  }\bibfield  {title} {\bibinfo {title} {Edge states and topological invariants
  of non-{H}ermitian systems},\ }\href
  {https://doi.org/10.1103/PhysRevLett.121.086803} {\bibfield  {journal}
  {\bibinfo  {journal} {Phys. Rev. Lett.}\ }\textbf {\bibinfo {volume} {121}},\
  \bibinfo {pages} {086803} (\bibinfo {year} {2018})}\BibitemShut {NoStop}%
\bibitem [{\citenamefont {Yokomizo}\ and\ \citenamefont
  {Murakami}(2019)}]{Yokomizo:2019PRL}%
  \BibitemOpen
  \bibfield  {author} {\bibinfo {author} {\bibfnamefont {K.}~\bibnamefont
  {Yokomizo}}\ and\ \bibinfo {author} {\bibfnamefont {S.}~\bibnamefont
  {Murakami}},\ }\bibfield  {title} {\bibinfo {title} {Non-{B}loch band theory
  of non-{H}ermitian systems},\ }\href
  {https://doi.org/10.1103/PhysRevLett.123.066404} {\bibfield  {journal}
  {\bibinfo  {journal} {Phys. Rev. Lett.}\ }\textbf {\bibinfo {volume} {123}},\
  \bibinfo {pages} {066404} (\bibinfo {year} {2019})}\BibitemShut {NoStop}%
\bibitem [{\citenamefont {Yokomizo}\ and\ \citenamefont
  {Murakami}(2023)}]{Yokomizo:2023PRB}%
  \BibitemOpen
  \bibfield  {author} {\bibinfo {author} {\bibfnamefont {K.}~\bibnamefont
  {Yokomizo}}\ and\ \bibinfo {author} {\bibfnamefont {S.}~\bibnamefont
  {Murakami}},\ }\bibfield  {title} {\bibinfo {title} {Non-{B}loch bands in
  two-dimensional non-{H}ermitian systems},\ }\href
  {https://doi.org/10.1103/PhysRevB.107.195112} {\bibfield  {journal} {\bibinfo
   {journal} {Phys. Rev. B}\ }\textbf {\bibinfo {volume} {107}},\ \bibinfo
  {pages} {195112} (\bibinfo {year} {2023})}\BibitemShut {NoStop}%
\bibitem [{\citenamefont {Yi}\ and\ \citenamefont {Yang}(2020)}]{Yi:2020PRL}%
  \BibitemOpen
  \bibfield  {author} {\bibinfo {author} {\bibfnamefont {Y.}~\bibnamefont
  {Yi}}\ and\ \bibinfo {author} {\bibfnamefont {Z.}~\bibnamefont {Yang}},\
  }\bibfield  {title} {\bibinfo {title} {Non-{H}ermitian skin modes induced by
  on-site dissipations and chiral tunneling effect},\ }\href
  {https://doi.org/10.1103/PhysRevLett.125.186802} {\bibfield  {journal}
  {\bibinfo  {journal} {Phys. Rev. Lett.}\ }\textbf {\bibinfo {volume} {125}},\
  \bibinfo {pages} {186802} (\bibinfo {year} {2020})}\BibitemShut {NoStop}%
\bibitem [{\citenamefont {Aharonov}\ and\ \citenamefont
  {Bohm}(1959)}]{Aharonov:1959PR}%
  \BibitemOpen
  \bibfield  {author} {\bibinfo {author} {\bibfnamefont {Y.}~\bibnamefont
  {Aharonov}}\ and\ \bibinfo {author} {\bibfnamefont {D.}~\bibnamefont
  {Bohm}},\ }\bibfield  {title} {\bibinfo {title} {Significance of
  electromagnetic potentials in the quantum theory},\ }\href
  {https://doi.org/10.1103/PhysRev.115.485} {\bibfield  {journal} {\bibinfo
  {journal} {Phys. Rev.}\ }\textbf {\bibinfo {volume} {115}},\ \bibinfo {pages}
  {485} (\bibinfo {year} {1959})}\BibitemShut {NoStop}%
\bibitem [{\citenamefont {Anandwade}\ \emph {et~al.}(2023)\citenamefont
  {Anandwade}, \citenamefont {Singhal}, \citenamefont {Paladugu}, \citenamefont
  {Martello}, \citenamefont {Castle}, \citenamefont {Agrawal}, \citenamefont
  {Carlson}, \citenamefont {Battle-McDonald}, \citenamefont {Ozawa},
  \citenamefont {Price},\ and\ \citenamefont {Gadway}}]{Anandwade:2023PRA}%
  \BibitemOpen
  \bibfield  {author} {\bibinfo {author} {\bibfnamefont {R.}~\bibnamefont
  {Anandwade}}, \bibinfo {author} {\bibfnamefont {Y.}~\bibnamefont {Singhal}},
  \bibinfo {author} {\bibfnamefont {S.~N.~M.}\ \bibnamefont {Paladugu}},
  \bibinfo {author} {\bibfnamefont {E.}~\bibnamefont {Martello}}, \bibinfo
  {author} {\bibfnamefont {M.}~\bibnamefont {Castle}}, \bibinfo {author}
  {\bibfnamefont {S.}~\bibnamefont {Agrawal}}, \bibinfo {author} {\bibfnamefont
  {E.}~\bibnamefont {Carlson}}, \bibinfo {author} {\bibfnamefont
  {C.}~\bibnamefont {Battle-McDonald}}, \bibinfo {author} {\bibfnamefont
  {T.}~\bibnamefont {Ozawa}}, \bibinfo {author} {\bibfnamefont {H.~M.}\
  \bibnamefont {Price}},\ and\ \bibinfo {author} {\bibfnamefont
  {B.}~\bibnamefont {Gadway}},\ }\bibfield  {title} {\bibinfo {title}
  {Synthetic mechanical lattices with synthetic interactions},\ }\href
  {https://doi.org/10.1103/PhysRevA.108.012221} {\bibfield  {journal} {\bibinfo
   {journal} {Phys. Rev. A}\ }\textbf {\bibinfo {volume} {108}},\ \bibinfo
  {pages} {012221} (\bibinfo {year} {2023})}\BibitemShut {NoStop}%
\bibitem [{\citenamefont {Singhal}\ \emph {et~al.}(2023)\citenamefont
  {Singhal}, \citenamefont {Martello}, \citenamefont {Agrawal}, \citenamefont
  {Ozawa}, \citenamefont {Price},\ and\ \citenamefont
  {Gadway}}]{Singhal:2023PRR}%
  \BibitemOpen
  \bibfield  {author} {\bibinfo {author} {\bibfnamefont {Y.}~\bibnamefont
  {Singhal}}, \bibinfo {author} {\bibfnamefont {E.}~\bibnamefont {Martello}},
  \bibinfo {author} {\bibfnamefont {S.}~\bibnamefont {Agrawal}}, \bibinfo
  {author} {\bibfnamefont {T.}~\bibnamefont {Ozawa}}, \bibinfo {author}
  {\bibfnamefont {H.}~\bibnamefont {Price}},\ and\ \bibinfo {author}
  {\bibfnamefont {B.}~\bibnamefont {Gadway}},\ }\bibfield  {title} {\bibinfo
  {title} {Measuring the adiabatic non-{H}ermitian berry phase in
  feedback-coupled oscillators},\ }\href
  {https://doi.org/10.1103/PhysRevResearch.5.L032026} {\bibfield  {journal}
  {\bibinfo  {journal} {Phys. Rev. Res.}\ }\textbf {\bibinfo {volume} {5}},\
  \bibinfo {pages} {L032026} (\bibinfo {year} {2023})}\BibitemShut {NoStop}%
\bibitem [{\citenamefont {Griffiths}\ and\ \citenamefont
  {Schroeter}(2018)}]{GriffitshBook}%
  \BibitemOpen
  \bibfield  {author} {\bibinfo {author} {\bibfnamefont {D.~J.}\ \bibnamefont
  {Griffiths}}\ and\ \bibinfo {author} {\bibfnamefont {D.~F.}\ \bibnamefont
  {Schroeter}},\ }\href@noop {} {\emph {\bibinfo {title} {{Introduction to
  Quantum Mechanics}}}},\ \bibinfo {edition} {3rd}\ ed.\ (\bibinfo  {publisher}
  {Cambridge University Press, Cambridge},\ \bibinfo {year} {2018})\BibitemShut
  {NoStop}%
\bibitem [{\citenamefont {Faugno}\ and\ \citenamefont
  {Ozawa}(2022)}]{Faugno:2022PRL}%
  \BibitemOpen
  \bibfield  {author} {\bibinfo {author} {\bibfnamefont {W.~N.}\ \bibnamefont
  {Faugno}}\ and\ \bibinfo {author} {\bibfnamefont {T.}~\bibnamefont {Ozawa}},\
  }\bibfield  {title} {\bibinfo {title} {Interaction-induced non-{H}ermitian
  topological phases from a dynamical gauge field},\ }\href
  {https://doi.org/10.1103/PhysRevLett.129.180401} {\bibfield  {journal}
  {\bibinfo  {journal} {Phys. Rev. Lett.}\ }\textbf {\bibinfo {volume} {129}},\
  \bibinfo {pages} {180401} (\bibinfo {year} {2022})}\BibitemShut {NoStop}%
\bibitem [{\citenamefont {Paiva}\ \emph {et~al.}(2023)\citenamefont {Paiva},
  \citenamefont {Aharonov}, \citenamefont {Tollaksen},\ and\ \citenamefont
  {Waegell}}]{Paiva:2023NJP}%
  \BibitemOpen
  \bibfield  {author} {\bibinfo {author} {\bibfnamefont {I.~L.}\ \bibnamefont
  {Paiva}}, \bibinfo {author} {\bibfnamefont {Y.}~\bibnamefont {Aharonov}},
  \bibinfo {author} {\bibfnamefont {J.}~\bibnamefont {Tollaksen}},\ and\
  \bibinfo {author} {\bibfnamefont {M.}~\bibnamefont {Waegell}},\ }\bibfield
  {title} {\bibinfo {title} {Aharonov--{B}ohm effect with an effective
  complex-valued vector potential},\ }\href
  {https://iopscience.iop.org/article/10.1088/1367-2630/acd4dd} {\bibfield
  {journal} {\bibinfo  {journal} {New Journal of Physics}\ }\textbf {\bibinfo
  {volume} {25}},\ \bibinfo {pages} {053017} (\bibinfo {year}
  {2023})}\BibitemShut {NoStop}%
\end{thebibliography}%
\end{document}